\newif\ifarxiv

\arxivtrue

\ifarxiv
\documentclass[screen,sigconf,nonacm,anonymous=false]{acmart}  
\else
\documentclass[screen,sigconf,anonymous=false]{acmart}  
\fi

 \newcommand{\implementationurl}{\url{https://github.com/znewman01/speranza}}
 \newcommand{\prurl}{\url{https://github.com/znewman01/fulcio/pull/1}}

\usepackage[binary-units,group-separator={,}]{siunitx}
\usepackage[capitalize]{cleveref}
\usepackage{enumitem}

\setlist[itemize,1]{leftmargin=0.4cm}

\newcommand{\Z}{\mathbb{Z}}
\newcommand{\hash}{\mathsf{H}}

\newcommand*{\secpar}{\ensuremath{\lambda}}
\newcommand*{\alg}[1]{\ensuremath{\mathsf{#1}}}
\newcommand*{\pp}{\ensuremath{\mathtt{pp}}}
\newcommand*{\sk}{\mathtt{sk}}
\newcommand*{\cert}{\mathtt{cert}}
\newcommand*{\skoidc}{\mathtt{sk}_\mathtt{oidc}}
\newcommand*{\skca}{\mathtt{sk}_\mathtt{ca}}
\newcommand*{\skcert}{\mathtt{sk}_\mathtt{cert}}
\newcommand*{\pk}{\mathtt{pk}}
\newcommand*{\pkoidc}{\mathtt{pk}_\mathtt{oidc}}
\newcommand*{\pkca}{\mathtt{pk}_\mathtt{ca}}
\newcommand*{\pkcert}{\mathtt{pk}_\mathtt{cert}}
\newcommand*{\cpub}{\mathtt{c}_\mathtt{pub}}
\newcommand*{\sub}{\mathtt{sub}}
\newcommand*{\id}{\mathtt{id}}
\newcommand*{\tok}{\mathtt{tok}}
\newcommand*{\yes}{\mathrm{yes}}
\newcommand*{\no}{\mathrm{no}}
\newcommand*{\adv}{\mathcal{A}}
\newcommand*{\advb}{\mathcal{B}}
\newcommand*{\negl}{\mathsf{negl}}

\newcommand*{\rgets}{\overset{\$}{\gets}}
\newcommand*{\simu}{\mathrm{Sim}}

\newcommand*{\aux}{\mathtt{aux}}
\newcommand*{\pkg}{\mathtt{pkg}}

\newcommand*{\ext}{\mathrm{Extract}}
\newcommand*{\ppp}{\pp_{\mathsf{pdrsn}}}
\newcommand*{\ppnizk}{\pp_{\mathsf{nizk}}}

\AtBeginDocument{%
  }

\copyrightyear{2023}
\acmYear{2023}
\setcopyright{rightsretained}
\acmConference[CCS '23]{Proceedings of the 2023 ACM SIGSAC Conference on Computer and Communications Security}{November 26--30, 2023}{Copenhagen, Denmark}
\acmBooktitle{Proceedings of the 2023 ACM SIGSAC Conference on Computer and Communications Security (CCS '23), November 26--30, 2023, Copenhagen, Denmark}\acmDOI{10.1145/3576915.3623200}
\acmISBN{979-8-4007-0050-7/23/11}

\ifarxiv
\else


\fi

\newcommand{\santiago}[1]{{\color{green} #1}}
\newcommand{\zjn}[1]{{\color{blue} #1}}

\newcommand{\karen}[1]{{\color{purple} #1}}

\newcommand{\ourthing}{Speranza}
\newcommand{\cocommitment}{identity co-commitment}
\newcommand{\cocommitments}{\cocommitment{}s}
\newcommand{\Cocommitment}{Identity co-commitment}
\newcommand{\Cocommitments}{\Cocommitment{}s}

\begin{document}

\title{\ourthing: Usable, privacy-friendly software signing}

\author{Kelsey Merrill}
\affiliation{%
  \institution{MIT CSAIL}
  \city{Cambridge, MA}
  \country{USA}}
\email{kmerrill@csail.mit.edu}

\author{Zachary Newman}
\affiliation{%
  \institution{Chainguard, Inc.}
  \city{Brooklyn, NY}
  \country{USA}}
\email{research@z.znewman.net}

\author{Santiago Torres-Arias}
\affiliation{%
  \institution{Purdue University}
  \city{West Lafayette, IN}
  \country{USA}}
\email{santiagotorres@purdue.edu}

\author{Karen R. Sollins}
\affiliation{%
  \institution{MIT CSAIL}
  \city{Cambridge, MA}
  \country{USA}}
\email{sollins@csail.mit.edu}
\renewcommand{\shortauthors}{Merrill et al.}

\begin{abstract}
Software repositories, used for wide-scale open software distribution, are a significant vector for security attacks.
Software signing provides authenticity, mitigating many such attacks.
Developer-managed signing keys pose usability challenges, but certificate-based systems introduce privacy problems.
This work, \ourthing{}, uses certificates to verify software authenticity but still provides anonymity to signers using zero-knowledge \emph{\cocommitments{}}.

In \ourthing{}, a signer uses an automated certificate authority (CA) to create a private identity-bound signature and proof of authorization.
Verifiers check that a signer was authorized to publish a package without learning the signer's identity.
The package repository privately records each package's authorized signers, but publishes only \emph{commitments} to identities in a public map.
Then, when issuing certificates, the CA issues the certificate to a distinct commitment to the same identity.
The signer then creates a zero-knowledge proof that these are \cocommitments{}.

We implemented a proof-of-concept for \ourthing{}.
We find that costs to maintainers (signing) and end users (verifying) are small (sub-millisecond), even for a repository with millions of packages.
Techniques inspired by recent key transparency systems reduce the bandwidth for serving authorization policies to \SI{2}{\kibi\byte}.
Server costs in this system are negligible.
Our evaluation finds that \ourthing{} is practical on the scale of the largest software repositories.

We also emphasize practicality and deployability in this project.
By building on existing technology and employing relatively simple and well-established cryptographic techniques, Speranza can be deployed for wide-scale use with only a few hundred lines of code and minimal changes to existing infrastructure.
Speranza is a practical way to bring privacy and authenticity together for more trustworthy open-source software.

\end{abstract}

\maketitle

\section{Introduction}
Open source software is of vital importance to today's society, as almost all codebases use it \cite{software_census}. 
However, the nature of open source software's development makes it especially hard to secure, due to both the sheer numbers of maintainers and their varying nature. 
Each of these maintainers' accounts, along with the repository itself, is a potential entry point for an attacker.

Attacks on software distribution via repositories can have enormous impact~\cite{node-event-stream, dependency-confusion}. In such attacks, malicious actors compromise software repository credentials~\cite{eslint}, repository infrastructure~\cite{github_issue}, or distribution channels~\cite{github_mitm_china} to publish malware masquerading as packages from maintainers that downstream developers implicitly trust.
This has led to tens of thousands of malicious downloads and hundreds of real-world incidents~\cite{os_ssc_attacks_review_ohm20}, including the United States government in the SolarWinds attack~\cite{solarwinds}, resulting in an Executive Order in May 2021 charging NIST with increasing the integrity of the software supply chain~\cite{EO2021}.

In response to these sorts of attacks, organizations such as Microsoft and the Center for Internet Security (CIS) recommend software signing as one tool for greater package repository security~\cite{cis-sscs, ms-sscs}. In a software signing scheme, maintainers digitally sign their software artifacts to assure users that these artifacts were actually produced by the expected maintainer. However, uptake has been low due to usability concerns \cite{digsigs_github}. To address the usability issues with classic digital signatures, \citet{newman2022sigstore} proposed Sigstore, which allows users to generate digital signatures without managing a key pair themselves, using their email addresses as identities rather than a long-lived key pair.

The privacy issue of exposing email addresses has limited uptake of Sigstore; proposals to add Sigstore signing to RubyGems~\cite{rubygems_rfc_2022} and Crates.io~\cite{rust_rfc_2021, rust_rfc_2023} have both stalled in part due to privacy concerns. npm has also expressed the importance of protecting maintainer personally identifiable information~\cite{npm_goals}.
Effectively protecting the vital resource that is open source software requires a software signing solution that is easy to use but still protects maintainer privacy.


\ourthing{} provides the same authenticity guarantees as the state of the art while hiding personally identifiable information from the public.
The package repository manages a public record of ownership for each package, dictating who is authorized to sign that package.
However, rather than storing a public key for each owner (as in a traditional package signing scheme) or the identity of each owner, such as an email address (as in a certificate-based scheme like Sigstore), \ourthing{} stores a \emph{commitment} to the identity of the owner (e.g., a Pedersen commitment~\cite{Pedersen_commits}).
Then, the certificate authority provides a certificate with a commitment to the identity of the requester, rather than the identity itself, as its subject.
The package maintainer then publishes a package, a signature over the package, a certificate containing the verification key for the package signature, and a zero-knowledge proof~\cite{ChaumPedersenProof} that the subject of the certificate and the package's public ownership record contain \emph{\cocommitments{}}---commitments to the same identity.

To summarize, the contributions of this work are:
\begin{itemize}
\item \emph{\Cocommitments{}}: a privacy-friendly technique for using third-party identity providers in security systems like package signing.
\item \ourthing{}, a system  for easy-but-private package signing:
  \begin{itemize}
    \item anonymized certificates with \cocommitments{}.
    \item techniques for efficiently managing an anonymized authorization record.
  \end{itemize}
\item An implemention and evaluation of \ourthing{}:
  \begin{itemize}
    \item signing and verifying microbenchmarks.
    \item a measure of server costs and bandwidth utilization.
    \item proof-of-concept implementation in Sigstore
  \end{itemize}
\item Analysis of data from PyPI on package ownership changes
\end{itemize}

\noindent
Like most privacy systems, \ourthing{} trades off between transparency,  privacy, usability, and performance.
Because certificates no longer store cleartext identities, or even consistent pseudonyms for cleartext identities, clients can no longer monitor for every time their identity is used.
Furthermore, \ourthing{} does not provide complete privacy, and maintainer identity is revealed to the certificate authority and the package repository.
Future work (\cref{sec:future-work}) might address these limitations.

In the remainder of the paper, we begin with background in \cref{sec:background} followed in \cref{sec:settingthreats} with enumeration of the elements of the underlying model, the goals of the attackers, and the goals of our system to mitigate those attacks.
With \cocommitments{} presented in \cref{sec:identity-co-commit}, we then discuss signing and verifying with them in \cref{sec:signing_verifying_flow}.
\Cref{sec:authorization_record} introduces the authorization record that reflects the authorization and policy for delegation to sign.
This is followed in \cref{sec:security_analysis} with a security analysis, in turn followed by our implementation and evaluation in \cref{sec:Implementation_eval}, further discussion and related work (\cref{sec:discussion_future_work}) and conclusions (\cref{sec:conclusion}).

\section{Background} \label{sec:background}

In this section, we present background on package repository security with an emphasis on digital signing, including schemes that allow signers to avoid managing long-lived key material.

A \emph{package repository} is a service for software distribution, allowing \emph{maintainers} to publish named \emph{packages}, and \emph{end users} to look up packages and download them. Almost all codebases (\SI{96}{\%} in a 2022 estimate~\cite{oss_risk_analysis_report_2023}) contain open-source software, often from package repositories like PyPI~\cite{pypi} and RubyGems~\cite{rubygems}.

Because package repositories are widely used, they are a valuable target for attackers, who can deliver malware to many end users at once.
This work focuses on attack distribution, where an end user requests a specific package and receives malware that the authorized maintainer of the package did not publish.
These attacks can happen for many reasons, including the compromise of a maintainer's credentials on the package repository or the compromise of repository itself. 



Digital software signing provides the basis for significant mitigation of this problem. A signature over a package allows a user to then verify that the given package was published as-is by a maintainer they trust. However, the introduction of cryptographic keys raises several problems, including key management and revocation in the case of compromised or lost keys \cite{whitten99, johnny-still-cant-encrypt}. One solution to developer-managed key systems is to create a public key infrastructure (PKI) linking digital identities to cryptographic keys via digital certificates.
The most popular deployment of PKI is the web PKI that secures communications between clients and web servers.
This allows clients to store a small root-of-trust (hundreds of entries) but communicate with web sites hosted under hundreds of millions of different domains.

The Sigstore project~\cite{newman2022sigstore} creates an automated ``PKI'' for software signing.
Maintainers run a command to sign their package, a browser window opens asking to authenticate with their email account, and the Sigstore infrastructure automatically issues a certificate for that email, creating a keypair linked to the email address, along with a timestamp.
When a user goes to download the artifact, they check that the signature matches the identity they are expecting, and they check the published certificate.
They also check that the signature was created while the certificate was valid.
Sigstore can support any digital identity whose provider supports OpenID Connect (OIDC) \cite{openid_2022}, an authentication protocol based on OAuth 2.0, such as a Google, Microsoft, or Facebook account.

Sigstore solves many of the usable security issues with traditional cryptographic key pairs.
By outsourcing authentication to an OIDC provider, developers only have to manage one account.
They also have account recovery provided by the OIDC provider. 
However, uptake of the Sigstore system has been limited in part due to privacy complaints from maintainers \cite{rubygems_rfc_2022, rust_rfc_2021, rust_rfc_2023, npm_goals}.
In order to use Sigstore, the developer's OIDC identity must be public.
For example, if using an email account, the developer's email address must be published.
Many developers wish to remain anonymous and find this unacceptable \cite{rust_rfc_2021, rust_rfc_2023, rubygems_rfc_2022}.
Providing a system that maintains developer privacy but is still usable would encourage the use of digital signatures on artifacts.
This is where \ourthing{} steps in.

Current software signing solutions also have the problem of ``key transparency'' concerning digital signatures. Classic digital signatures rely on the assumption that end users can securely learn the public verification keys associated with a particular package. In practice, this is nontrivial, because the repository itself is not a reliable source of those keys. If end users simply query the package repository for the key associated with a package, this provides little extra protection on top of the checks a repository performs on package publication.

The Update Framework (TUF)~\cite{tuf} systematically addresses many of these issues by separating different responsibilities into different \emph{roles} with corresponding keys, and then recording the mapping between the roles and packages as a large list the user can download.
Keys for important roles can be stored offline, reducing the risk of compromise and allowing recovery from the compromise of online keys.
These roles support \emph{thresholds} for greater security.
In subsequent work, Diplomat~\cite{diplomat} introduces \emph{delegations} and a scheme for locking in delegations such that even the package repository itself cannot reverse them. 

In TUF, forking/equivocation/split-view attacks are still possible; a TUF repository can present different views to different end users.
Further, an attacker who compromises a repository could ``rewind'' to undo evidence of their attack.
To help detect compromise, transparency logs~\cite{ct}, a technique originating in the web public key infrastructure, require all operations to be posted to a public, tamper-proof log.
The log can be ``audited'' by third party \emph{monitors} which communicate amongst themselves~\cite{meiklejohn_gossip,parakeet}. This in turn allows for the detection of both split-view attacks and general unauthorized activity that might indicate a compromise.
\ourthing{}, originally proposed in \cite{kmerrill-thesis}, brings this work together and formalizes it in a state machine model to provide a solution to the key transparency problem.

\section{Setting and Threat Model} \label{sec:settingthreats}

With this background in mind, we consider our setting and model. We first introduce the parties involved in the system, and then consider the nature of the attackers. We then consider our system goals, as well as some key assumptions.

\subsection{Parties and Roles}
The system has the following parties and roles:
\begin{itemize}
    \item \textbf{Signers}: individuals wishing to sign an artifact
    \item \textbf{Verifiers}: individuals wishing to verify the authenticity of an artifact.
    \item \textbf{Package repository}: independent service hosting artifacts for download. The repository also maintains an \textit{authorization record}, a data structure managed by the package repository that maps artifacts to those identities with the authorization to sign them.
    \item \textbf{Identity provider}: entity vouching that an individual controls an identity.
    \item \textbf{Certificate authority}: entity that verifies identity tokens and issues certificates to signers.
    \item \textbf{Monitors}: third parties auditing the package repository for correctness and consistency.
\end{itemize}

\noindent
For example: Alice (signer) wishes to sign artifact \texttt{Foo} attesting to its authenticity.
She generates a key pair, authenticates with Google (identity provider), and receives a certificate from the certificate authority. She then uses her key pair to sign \texttt{Foo}. 
Bob (verifier) wishes to download \texttt{Foo} and check its authenticity.
He verifies that the signature on \texttt{Foo} is valid and the identity in the certificate is the same as the identity stored under \texttt{Foo} in the authorization record.


\subsection{Attacker Capabilities and Goals} \label{sec:attacker_capabilities}
We assume that an attacker may compromise the maintainer's package repository account. We examine how security in our system degrades under further compromise of the package repository or the certificate authority in \cref{sec:system_compromise}.
Maintainers' identity provider passwords or cryptographic material belonging to others, such as secret keys, are assumed to be inaccessible to attackers. We also assume a one-time trusted setup phase, perhaps conducted by a standards-setting body, that no attacker may compromise. An attacker may have the following goals:

\subsubsection{Get users to run malicious code} An attacker may try to get unsuspecting users to download and run malicious code. They might do this by modifying an existing package to include their code or replacing an existing package with their code.
The ability to run arbitrary code on someone else's system enables an attacker to achieve any number of goals including using computing resources, introducing malware into the user's system, or introducing security vulnerabilities for later exploit.

\subsubsection{Rollback attacks} An attacker may try to change a package back to a previous version with known security vulnerabilities. This allows the package to largely function as usual and likely remain undetected by the user but still introduces security vulnerabilities that the attacker can use for later exploit.

\subsubsection{Violate the privacy of maintainers} An attacker may try to learn the identity of the maintainer of a given package, determine that the same maintainer is publishing multiple given packages, or confirm or deny a guess about the maintainer of a given package. A maintainer's OIDC identity likely includes personally identifiable contact information. An attacker could use this contact information to contact the maintainer themselves to try to influence the maintainer socially, conduct marketing campaigns, or engage in harassment. An attacker could also collect many such identities and attempt to sell this personally identifiable information for advertising purposes.

\subsection{System Goals} \label{sys-goals}
We have the following goals for \ourthing{}: correctness, authenticity, privacy, transparency, and deployability.
We introduce security goals informally here, and we define these more formally in \cref{sec:formal-defs}.

\subsubsection{Correctness}
A signature generated by an honest signer should always be accepted by the verifier.

Correctness means that the system has no ``false positives'' when signalling malicious behavior.
When honest parties are using \ourthing{}, the user should be able to download artifacts successfully.

\subsubsection{Authenticity}
Verifiers should be able to confirm that an authorized signer (as defined by the authorization record) attested to the validity of the artifact.

Authenticity means that \ourthing{} is secure from the user's point of view. When users download artifacts using \ourthing{}, they are convinced that the artifact is valid according to the rightful maintainer of that artifact.

\subsubsection{Privacy}
Only the signer, the certificate authority, and the package repository should know the identity of the signer. Verifiers and other signers should not be able to determine anything about a given signer's  identity. They should not learn the identity, learn which artifacts are signed by the same identity, or have the ability to confirm or deny guesses about a signer's identity.

Privacy means that \ourthing{} is secure from the maintainer's point of view. It means that maintainer identities will not be made public when publishing artifacts with \ourthing{}.

Note that \ourthing{} aims to provide privacy from \textit{the public}, not \textit{the system}.
Under this privacy goal, maintainer identities may be exposed to the package repository and the certificate authority.
Also note that this privacy goal allows for maintainer linkages across the \emph{same} package, but not across different packages. A user can see that some maintainer $x$ signed package \texttt{Foo} multiple times, but they cannot tell if maintainer $x$ is also signing package \texttt{Bar}.

\subsubsection{Transparency}
All changes to the authorization record should be public and verifiable. Thus, malicious tampering with the authorization record should be detectable.

"Tampering" can be divided into two categories - consistency and correctness. Consistency means that the server cannot equivocate about the authorization record, and all clients see the same record. Correctness is the idea that all mappings in the authorization record are correct and as they should be, i.e., there are no mappings between a package and a non-owner of that package (like an adversary). Our transparency goal requires that both types of tampering be detectable.

Transparency means that we can respond to compromise of the authorization record. If compromise can be detected, users, maintainers, and other parties can then respond to that event, work to recover from compromise, and use alternate methods in the meantime to protect themselves.


This transparency goal excludes the identity provider, the certificate authority, and the repository as a whole from compromise detection. These tradeoffs in transparency are made in order to gain privacy. More discussion of the privacy-transparency trade-off inherent in this system can be found in \cref{sec:discussion-transparency}.

\subsubsection{Deployability}
This system should be easily usable and deployable with minimal changes to existing infrastructure.

\subsection{Cryptographic Assumptions}
We assume the existence of a collision-resistant hash function (CRHF) and a commitment scheme with efficient zero-knowledge arguments of commitment equality (see \cref{sec:identity-co-commit}); we realize such a scheme using the Chaum-Pedersen construction~\cite{ChaumPedersenProof} in a group where the discrete logarithm problem is assumed hard.
We also require a digital signature scheme.
Specifically, we use SHA2-512 as a CRHF, the Ristretto255~\cite{ristretto} group over Curve25519~\cite{curve25519}, and Ed25519 signatures.

\section{Identity Co-Commitments} \label{sec:identity-co-commit}
Identity co-commitments incorporate identity into a system while respecting privacy.
This section describes identity co-commitments and the cryptographic techniques they use.

\subsection{Cryptographic Commitments}
\label{sec:commitments}
Identity co-commitments rely on a cryptographic commitment scheme and zero-knowledge proof of commitment equality. 
Cryptographic commitments have the following algorithms:

\begin{itemize}
  \item $\alg{Generate}(\secpar) \to \pp$: generate public parameters.
  \item $\alg{Commit}(\pp, m) \to (c, r)$: create commitment $c$ to message $m$ with randomness $r$.
  \item $\alg{Verify}(\pp, m, c, r) \to \yes / \no$: check the commitment.
\end{itemize}

\noindent
They satisfy three security properties 
\ifarxiv
(formal definitions can be found in \cref{sec:crypto_appendix})
\fi
:
\begin{itemize}
    \item Correctness: a commitment to some value $x$ can always be opened to value $x$.
    \item Hiding: a commitment cannot be ``inverted'' to reveal the pre-image without the random key needed to ``unlock'' it.
    \item Binding: a commitment to some value $x$ cannot be opened to a different value $x'$.
\end{itemize}

\noindent
A proof of commitment equality is a zero-knowledge proof-of-knowledge convincing a verifier that (a) these two commitments are commitments to the same message and (b) the prover knows what that message is using the following algorithms:

\begin{itemize}
  \item $\alg{GenerateEq}(\secpar) \to \pp$: generate public parameters for proofs of equality.
  \item $\alg{ProveEq}(\pp, m, c_1, r_1, c_2, r_2) \to \pi \text{ or } \bot$: prove $c_1$ and $c_2$ are commitments to the same identity with keys $r_1$ and $r_2$ respectively.
  \item $\alg{VerifyEq}(\pp, c_1, c_2, \pi) \to \yes / \no$: check an equality proof.
\end{itemize}

\noindent
They have three security properites:
\begin{itemize}
    \item Completeness: the verifier will always accept a proof from an honest prover.
    \item Zero knowledge: the verifier cannot learn anything about the value committed to from the proof.
    \item Knowledge soundness: in order for the verifier to accept the proof, the prover must know the value committed to.
\end{itemize}

\noindent
We construct identity co-commitments using Pedersen commitments and Chaum-Pedersen proofs of commitment equality 
\ifarxiv
(constructions detailed in \cref{sec:crypto_appendix}).
\fi
However, this technique does not actually require Pedersen commitments specifically. It only requires a commitment scheme with the following properties:
\begin{itemize}
    \item hiding and binding (all commitment schemes)
    \item efficient zero-knowledge proof of commitment equality
\end{itemize}
\noindent
Pedersen commitments have a particularly simple proof of commitment equality, making them a good choice for this scheme.

\subsection{Identity Co-Commitments}
Identity co-commitments are useful when pseudonyms are desirable, but long-term pseudonyms are not sufficient.
Long-term pseudonyms (like verifiable credentials \cite{verifiable-credentials}) allow someone to be linked across uses of that pseudonym.
Even if the underlying identity is hidden, it is known that one identity is doing all the actions associated with that given pseudonym.
Identity co-commitments provide \textit{selectively linkable} pseudonyms.
They allow one party to link pairs of signatures under their identity together when and how they wish.

\subsubsection{Parties and roles}
Identity co-commitments have the following roles:
\begin{itemize}
    \item \textbf{Prover}: an individual holding a given identity that wishes to keep that identity private.
    \item \textbf{Verifier}: an individual wishing to verify a given identity.
    \item \textbf{Authority}: a third party that knows the identity of the prover and keeps a public and private record of this identity.
\end{itemize}

\subsubsection{Methods}
Identity co-commitments have the following methods, implemented in terms of Pedersen commitments:
\setlist[enumerate,1]{leftmargin=0.7cm}
\begin{itemize}
  \item $\alg{CoCo.Generate}(\secpar) \to \pp = (\ppp, \ppnizk)$: create public parameters for co-commitments.
  \begin{enumerate}
      \item Return $(\alg{Pedersen.Generate}(\secpar), \alg{Pedersen.GenerateEq}(\secpar))$.
  \end{enumerate}
      \item $\alg{CoCo.Commit}(\pp, \id) \to (c, r)$: create a commitment.
  \begin{enumerate}
      \item Return $\alg{Pedersen.Commit}(\ppp, \id)$
  \end{enumerate}
  \item $\alg{CoCo.Prove}(\pp, \id, c_1, r_1, c_2, r_2) \to \pi \text{ or } \bot$: prove $c_1$ and $c_2$ are co-commitments.
  \begin{enumerate}
      \item Return $\alg{Pedersen.ProveEq}(\ppp, \ppnizk, \id, c_1, r_1, c_2, r_2)$
  \end{enumerate}
  \item $\alg{CoCo.Verify}(\pp, \id, c, r) \to \yes / \no$
  \begin{enumerate}
      \item Return $\alg{Pedersen.Verify}(\ppp, \id, c, r)$
  \end{enumerate}
  \item $\alg{CoCo.VerifyEq}(\pp, c_1, c_2, \pi) \to \yes / \no$
  \begin{enumerate}
      \item Return $\alg{Pedersen.VerifyEq}(\ppp, \ppnizk, c_1, c_2, \pi)$
  \end{enumerate}
\end{itemize}

\noindent
We also define $\alg{FillGraph}$, which creates a ``co-commitment graph'': a graph where nodes are commitments to the same identity, and edges are proofs of equality between them.
\begin{itemize}
    \item $\alg{FillGraph}(\pp, G, m) \to H$: for graph structure $G$ and message $m$, returns an instantiated graph $H$ where nodes are commitments to $m$ computed as $\alg{Pedersen.Commit}(\ppp, m)$, and edges are equality proofs between their connected nodes computed as $e_{ij} = \alg{Pedersen.ProveEq}(\pp, m, c_i, r_i, c_j, r_j)$.
\end{itemize}

\subsubsection{Security properties}
Identity co-commitments are a particular deployment of regular cryptographic commitments. Thus, they retain all of the security properties of Pedersen commitments discussed in \cref{sec:commitments}, and they have the following additional properties:
\begin{itemize}
    \item \textbf{Privacy of Identity}: co-commitments do not leak the underlying identity.
    
    For all $\id_0, \id_1, G$, where $G$ is an undirected graph structure: \\
    \begin{align*}
        (\pp, \alg{FillGraph}(\pp, G,\id_0)) \approx_c (\pp, \alg{FillGraph}(\pp, G,\id_1))
    \end{align*}
    (where $\approx_c$ indicates computational indistinguishability).

    \item \textbf{Linkability}: if two commitments are co-commitments, even transitively, then they are commitments to the same identity.

    For all connected graphs $H$, and all PPT adversaries $\adv$:
    \[
        \Pr\left[
        \begin{array}{l}
            \pp \gets \alg{CoCo.Generate}(\secpar) \\
            (H, \{ \id_i \}, \{ r_i \}) \gets \adv(\pp) \\
            \text{such that  } |\{ \id_i \}| > 1 \text{ and } \\
            \forall e_{ij}, \alg{CoCo.VerifyEq}(\pp, c_i, c_j, e_{ij}) = \yes \text{ and } \\
            \forall \id_i, c_i, r_i, \alg{CoCo.Verify}(\pp, \id_i, c_i, r_i) = \yes
        \end{array}
        \right]
        \leq \negl(\secpar).
    \]

    This means that no adversary should be able to generate a connected graph of co-commitments such that all of the following hold: \begin{itemize}
        \item The set of nodes contain commitments to at least two distinct identities ($\{ \id_i \}| > 1$)
        \item All of the commitment equality proofs on the edges verify
        \item All of the commitments on the nodes verify
    \end{itemize}
\end{itemize}

\subsection{Construction}
Identity co-commitments require the following setup phase:
\begin{enumerate}
    \item Trusted setup: $\pp = \alg{CoCo.Generate}(\secpar)$ is securely computed and published.
    \item Authority computes $c_a, r_a = \alg{CoCo.Commit}(\pp, \id)$. They publish $c_a$ in the public record, and they store $(\id, r_a)$ in the private record.
\end{enumerate}

\noindent
When a prover wishes to use her identity, they take the following steps:
\begin{enumerate}
    \item Prover computes $c_p, r_p = \alg{CoCo.Commit}(\pp, \id)$.
    \item Prover communicates with the Authority to retrieve $r_a$. This should involve some form of authentication.
    \item Prover computes $\pi = \alg{CoCo.Prove}(\pp, \id, c_a, r_a, c_p, r_p)$.
    \item Prover publishes $c_p, \pi$.
\end{enumerate}

\noindent
When a verifier wishes to check the identity co-commitment, they compute $\alg{CoCo.Verify}(\pp, c_a, c_p, \pi)$.

\subsection{Co-Commitments Security Analysis}
The following provides intuition about why identity co-commitments provide privacy of identity and linkability as defined above.
\ifarxiv
Full proofs can be found in \cref{sec:coco-proofs-appendix}
\fi

\subsubsection{Privacy of identity}
Intuitively, privacy of identity must hold by the non-interactive zero knowledge property and the hiding property of commitments. A graph is made up of commitments for nodes and proofs for edges. Hiding says that the commitments do not reveal any information about the underlying identity, and NIZK says that the proofs do not either.

The proof is a hybrid argument. Starting with a graph of $\id_0$, the edges can one at a time be switched out for simulated proofs without an adversary being able to distinguish by the NIZK property. Then, the graph can be replaced with commitments to $\id_1$ and the relevant simulated proofs by the hiding property. The process can then go in reverse, with simulated edges being replaced with real proofs by the NIZK property, until we arrive at a graph of $\id_1$.

\subsubsection{Linkability}
Intuitively, linkability must hold by the knowledge soundness property and the binding property of commitments. If a connected graph contains a commitment to a different identity than the nodes it is connected to, then either the proof edges are not sound, or an adversary was able to open a commitment to something other than the identity it was computed with.

The proof is a reduction to the binding game that also uses the extractor from the knowledge soundness definition. If there exists some adversary that can win the linkability game with non-negligible probability, there must also exist an efficient extractor. The extractor can then be used to win the binding game.

\section{Speranza Protocols} \label{sec:signing_verifying_flow}

This section contains a walkthrough of the three protocols that make up \ourthing{}: package registration, signing, and verifying. The system builds on Sigstore \cite{newman2022sigstore}, and many of the components are similar or identical to that system.
We leave the authorization record as a generic mapping data structure from package to identity.
More details on our proposed implementations for the authorization record follow in \cref{sec:authorization_record}.

\subsection{Preliminaries}
This section contains APIs for other system components outside of \cocommitments{} that we will reference below.

\paragraph{Digital signatures}
We use standard asymmetric digital signature algorithms:
\begin{itemize}
    \item $\alg{DigSig.Generate}(\secpar) \to (\sk, \pk)$: generate key pair.
    \item $\alg{DigSig.Sign}(\sk, m) \to \sigma$: sign message $m$ with secret key $\sk$.
    \item $\alg{DigSig.Verify}(\pk, m, \sigma) \to \yes / \no$: check that $\sigma$ is a valid signature over message $m$ for public key $\pk$.
\end{itemize}
Security properties for digital signatures can be found in 
\ifarxiv
\cref{sec:crypto_appendix}.
\fi

\paragraph{X.509 certificates}
We use the following algorithms for creating and verifying signing certificates (for simplicity, ignoring certificate details like the public key and validity periods):
\begin{itemize}
    \item $\alg{X509.Generate}(\secpar) \to (\skcert, \pkcert)$: generate key pair.
    \item $\alg{X509.SignCert}(\skcert, \sub, \pk) \to \cert$: sign $\sub$ and $\pk$ with $\skcert$ to produce $\cert$.
    \item $\alg{X509.VerifyCert}(\pkcert, \cert) \to \sub \text{ or } \bot$: verify that $\cert$ is valid according to $\pkcert$.
\end{itemize}

\paragraph{OIDC}
We use a simplified model of an OpenID Connect identity provider:
\begin{itemize}
    \item $\alg{OIDC.Generate} \to (\skoidc, \pkoidc)$: generate key pair.
    \item $\alg{OIDC.Issue}(\skoidc, \id) \to \tok$: issue $\tok$ to $\id$.
    \item $\alg{OIDC.Verify}(\pkoidc, \tok) \to \id \text{ or } \bot$: verify that $\tok$ is valid for $\pkoidc$, and return the underlying $\id$ if so.
\end{itemize}

\paragraph{Authorization record}
The authorization record maps from packages to owners:
\begin{itemize}
    \item $\alg{AuthRecord.Register}(\pkg, \id)$: register $\pkg$ to $\id$ in the authorization record
    \item $\alg{AuthRecord.Lookup}(\pkg) \to c$: look up $\pkg$ in the authorization record, and return a commitment $c$ to the identity that owns it.
\end{itemize}

\paragraph{Certificate authority}
We use a simplified model of an OIDC-backed certificate authority:
\begin{itemize}
\item $\alg{CA.Generate}(\secpar) \to (\skca, \pkca)$: generate key pair.
\item $\alg{CA.Issue}(\skca, \pkoidc, \tok) \to \cert$: Issues a certificate to the the identity corresponding to token. Puts a commitment to that identity in the subject instead of the identity itself. See steps below:
   \begin{enumerate}
   \item $\id \gets \alg{OIDC.Verify}(\pkoidc, \tok)$
   \item If $\id = \bot$ abort
   \item $c, r \gets \alg{CoCo.Commit}(\id)$
   \item $\cert \gets \alg{X509.Issue}(\sk, c)$
   \item Return $(\cert, r)$
   \end{enumerate}
\end{itemize}

\subsection{Setup}
Before any of the \ourthing{} protocols can be run, the trusted setup phase for Pedersen commitments and non-interactive zero knowledge proofs must be run.
This means securely computing $\pp = \alg{CoCo.Generate}(\secpar)$, then publishing $\pp$.

\subsection{Package Registration}
Before a maintainer can sign a given package, they must first register that package with the repository. Registration contains the following steps (see \cref{fig:registration}):
\begin{figure}
    \centering
    \includegraphics[width=\linewidth]{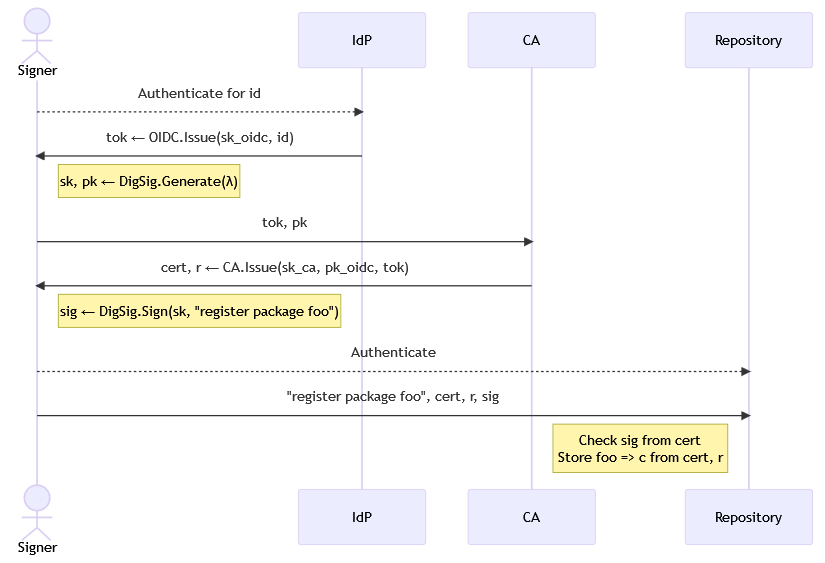}
    \caption{Speranza package registration protocol}
    \label{fig:registration}
\end{figure}

\begin{enumerate}
    \item Maintainer authenticates with the Identity Provider.
    \item Maintainer receives an OIDC token from the Identity Provider.
    \item Maintainer generates a Digital Signature key pair.
    \item Maintainer requests a certificate from the CA; sends their public key and their OIDC token to the CA.
    \item CA issues a certificate; returns the certificate and the commitment key to the signer.
    \item Maintainer uses their key pair to sign a message requesting that package \texttt{foo} be registered.
    \item Maintainer authenticates with the package repository (logs in).
    \item Maintainer sends message, certificate, commitment key, and signature to the package repository.
    \item Package repository checks that the certificate is valid and that the signature over the message is valid. If so, the repository registers package \texttt{foo}. The repository privately stores a mapping from \texttt{foo} to the commitment key, and it publicly stores a mapping from \texttt{foo} to the commitment in the subject of the certificate (see $\alg{CA.Issue}$).
\end{enumerate}

\subsection{Signing}
\begin{figure}
    \centering
    \includegraphics[width=\linewidth]{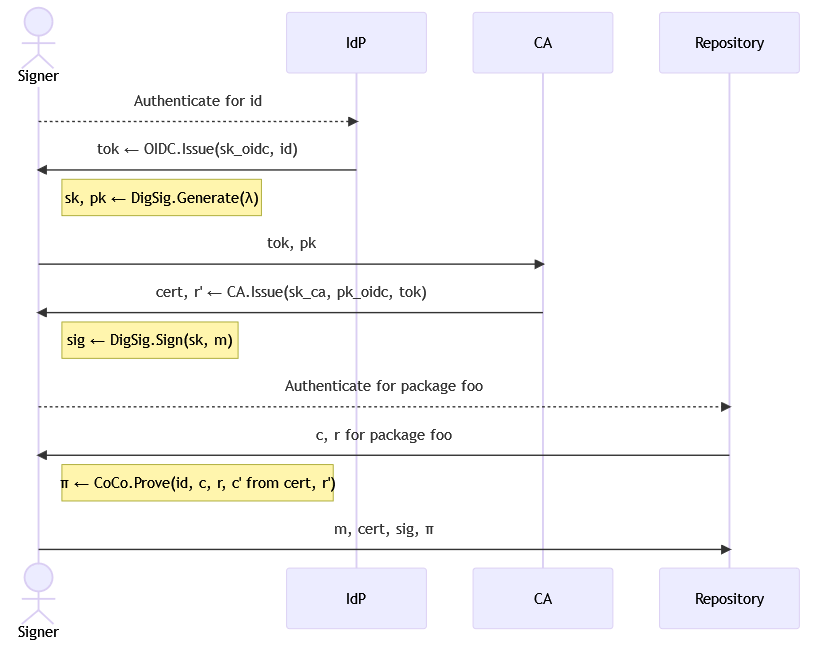}
    \caption{Speranza signing protocol}
    \label{fig:signing}
\end{figure}

Once a maintainer wishes to sign an artifact, they run the following protocol (see \cref{fig:signing}):
\begin{enumerate}
    \item Signer authenticates with the Identity Provider.
    \item Signer receives an OIDC token from the Identity Provider.
    \item Signer generates a Digital Signature key pair.
    \item Signer requests a certificate from the CA; sends their public key and their OIDC token to the CA.
    \item CA issues a certificate; returns the certificate and the commitment key to the signer.
    \item Signer uses their key pair to sign the package.
    \item Signer authenticates with the repository.
    \item Signer receives the repository's stored commitment and commitment key for the package.
    \item Signer computes an equality proof for the commitment on the certificate and the commitment returned from the repository (creating an identity co-commitment).
    \item Signer sends package, certificate, signature over the package, and the equality proof to the repository for publishing.
    \item Signer can destroy their ephemeral key pair and commitment keys.
\end{enumerate}

\subsection{Verification}
\begin{enumerate}
    \item Verifier retrieves the package, the certificate, the signature over the package, and the equality proof from the package repository.
    \item Verifier retrieves the commitment associated with the package from the authorization record.
    \item Verifier checks that the certificate is valid; if not, reject.
    \item Verifier checks that the signature over the package is valid; if not, reject.
    \item Verifier checks that the \cocommitments{} is valid; if not, reject. If so, accept.
\end{enumerate}

Once all of the relevant pieces have been retrieved from the package repository, verifying can be thought of as an algorithm: \\
$\alg{Speranza.Verify}(\pkca, \pp, \pkg, \sigma, \cert, \pi)$: \begin{enumerate}
    \item $\sub \gets \alg{X509.VerifyCert}(\pkca, \cert)$; If $sub = \bot$, return $\no$
    \item $\cpub \gets \alg{AuthRecord.Lookup}(\pkg)$
    \item If not $\alg{DigSig.Verify}(\cert.\pk, \pkg, \sigma)$, return $\no$
    \item Return $\alg{CoCo.Verify}(\pp, \cpub, \sub, \pi)$
\end{enumerate}


\section{Authorization Record}
\label{sec:authorization_record}
Up to this point, we have assumed a trusted authorization record.
We have also not explained how the server is prevented from misrepresenting the state of authorization to users or equivocating on artifact record,
nor have we described how the map can represent more complicated ownership, delegation, and publishing policies than a one-to-one artifact-signer mapping.
In this section, we explore different models of trust and methods for preventing the server from acting dishonestly.
We also provide methods for implementing complex delegation policies, such as those present in TUF \cite{tuf}, as well as methods for greater efficiency for clients using the authorization record.

For clarity, we first present the authorization record without any concern for privacy.
In \cref{sec:anon_signer_map}, we describe how the same techniques from the signature scheme in \cref{sec:signing_verifying_flow} can be used to anonymize it.

\subsection{State Machine Model}
\label{sec:state-machine}
We begin by introducing a model of an authorization record as a state machine, with allowed and disallowed transitions. Note that there is no notion of "security" in this section. The authorization record is assumed to be held in entirety by all parties.

\subsubsection{Basic Authorization Record}
An authorization record must support the following algorithms:

\begin{itemize}
    \item $\alg{Initialize}(\alg{artifacts}, \alg{signers}) \rightarrow \alg{AuthRecord}$:
    \\takes in a list of artifacts and a corresponding list of signers and returns an authorization record.
    \item $\alg{Lookup}(\alg{artifact}) \rightarrow \alg{Signer}$: 
    \\takes in an artifact and returns the signer associated with that artifact.
    \item $\alg{Register}(\alg{artifact}, \alg{signer})$: sets the signer for a given artifact (without authorization).
    \item $\alg{Update}(\alg{artifact}, \alg{new\_signer}, \alg{proof})$: takes in an existing artifact and the new signer as well as a proof that the update is authorized (for instance, a digital signature by the previous signer over the new signer). The transition only occurs if the proof verifies.
\end{itemize}

If there are no concerns for bandwidth efficiency, the authorization record can be backed by a simple hashmap of $(\alg{artifact}, \alg{signer})$ mappings.
However, this requires that clients download the entire authorization record. In order to ensure that state changes the server makes are honest (calls to $\alg{Update}$), the client will need to replicate all state changes in their own copy of the authorization record and also verify any calls to $\alg{Update}$ by checking the proof themselves. This is highly inefficient.

Still lacking from this model is support for more complicated ownership policies. The following sections will discuss support for these policies as well as techniques for greater client efficiency.

\subsubsection{Ownership policies}
In order to implement more complex ownership policies, we introduce an object, $\alg{Policy}$, for describing ownership, delegation, and publishing rules. A policy needs the following algorithms:
\begin{itemize}
    \item $\alg{CheckPublish}(\alg{artifact}, \alg{proof}) \rightarrow \yes / \no$:
    checks whether the artifact is authorized, per this policy. The type of $\alg{proof}$ varies by policy.
    \item $\alg{CheckPolicyChange}(\alg{update}, \alg{proof}) \rightarrow \yes / \no$:
    checks whether a requested change to this policy is authorized. The types of $\alg{update}$ and $\alg{proof}$ vary by policy.
\end{itemize}

Algorithms $\alg{CheckPublish}$ and $\alg{CheckPolicyChange}$ are both public and computable by the client as well as the server.
Transitions in the authorization record state machine are now:
\begin{itemize}
    \item $\alg{Initialize}(\alg{artifacts}, \alg{policies}) \rightarrow \alg{AuthRecord}$
    \item $\alg{Lookup}(\alg{artifact}) \rightarrow \alg{Policy}$
    \item $\alg{Register}(\alg{artifact}, \alg{policy})$
    \item $\alg{Update}(\alg{artifact}, \alg{new\_policy}, \alg{proof})$
\end{itemize}

Note that all instances of signer identities have been replaced with policies. A policy contains relevant signer identities.



\paragraph{Remaining goals}
With this construction of the authorization record (dictionary mapping \alg{(artifact, policy)}), the authorization record now supports arbitrary artifact ownership structures and changes in the artifacts themselves as well as in ownership. The remaining issue is client efficiency. In this model, the client has to download the entire authorization record. In the following sections, we explore techniques for making these operations more bandwidth-efficient for the client while also providing authenticity guarantees.

\subsection{Compressing the Authorization Record for Users: Third-Party Monitoring}  \label{sec:simple-merkle-and-auditing}
Instead of sending the entire authorization record to users, the repository can use a Merkle tree and send only a \emph{digest} of the authorization record.
Specifically, a basic authorization record is a dictionary from artifacts to policies.
We can use a Merkle prefix tree (as in~\cite{coniks}) to store the record.
The server publishes the entire Merkle tree in the clear, and it also publishes the Merkle digest (constant-sized---typically around \SI{32}{\byte}).
With this structure, the client only needs to download the constant-sized digest and a logarithmic-sized ``lookup proof'' (\SI{1.5}{\kibi\byte} for 1 million packages) to verify that the server has acted honestly on a lookup call.

We rely on third-party monitors to verify both the correctness and consistency.
A monitor can replay \emph{each} change to the tree, verify its correctness according to the state machine model (\cref{sec:state-machine}), and sign the new digest.
If a client sees signatures from the monitors for a digest, they know that the monitors have checked the record for that digest.

This model now has a notion of security concerning the consistency of the server's responses to $\alg{Lookup}$ calls with the true authorization record. This security is acheived from the security properties of a Merkle tree. When the Merkle digest is published, the server is \textit{bound} to a particular version of the authorization record, and all $\alg{Lookup}$ responses must be consistent with this one version or they will not verify.

\subsection{Anonymizing the Authorization Record} \label{sec:anon_signer_map}
So far in this section, we have ignored the privacy concerns motivating our proposed system. All signer privacy is lost if signer identities are published in the clear in the authorization record. This subsection will discuss how the same techniques used in signature scheme described in \cref{sec:signing_verifying_flow} can be used to anonymize the authorization record.

Recall that our current model of the authorization record has the following algorithms:
\begin{itemize}
    \item $\alg{Initialize}(\alg{artifacts}, \alg{policies}) \to \alg{AuthRecord}$
    \item $\alg{Lookup}(\alg{artifact}) \to (\alg{Policy}, \pi)$
    \item $\alg{Register}(\alg{artifact}, \alg{policy})$
    \item $\alg{Update}(\alg{artifact}, \alg{new\_policy}, \alg{proof})$
\end{itemize}

In this model, signer identities could be exposed within policies. Recall that a policy has the following algorithms:
\begin{itemize}
    \item $\alg{CheckPublish}(\alg{artifact}, \alg{proof}) \rightarrow \yes \text{ or } \no$:
    \item $\alg{CheckPolicyChange}(\alg{update}, \alg{proof}) \rightarrow \yes \text{ or } \no$:
\end{itemize}

Also recall that these are public algorithms, and any state contained in a $\alg{Policy}$ is also public. This state likely includes a list or tree of signer identities that are authorized to publish a artifact of make changes to its policy. The $\alg{proof}$ input likely includes signatures providing authenticity with publish or policy change requests.

As in \cref{sec:signing_verifying_flow}, any place where a signer identity would be used can instead be replaced with a commitment to that identity to provide privacy. In the same way, a commitment equality proof can be used to prove a match with the commitments stored inside a policy. The signature scheme described in \cref{sec:signing_verifying_flow} can then be used out of the box to provide signatures for the $\alg{proof}$ input.

Concrete examples of how verifying a publish event or changing a artifact policy might look with an anonymized authorization record can be found in \cref{sec:example-appendix}.

\section{Security Analysis}
\label{sec:security_analysis}
In this section, we first more formally define the security properties of \ourthing{} introduced in \cref{sys-goals}.
We then discuss how \ourthing{} meets these security properties.
Next, we discuss how these security properties protect against attacker goals in an honest system.
Lastly, we describe how the security of \ourthing{} degrades under system compromise.

\subsection{Formalizing Security Definitions}
\label{sec:formal-defs}
\paragraph{Correctness}
For all $\pkg, \id$, 
\[
    \Pr\left[
    \begin{array}{l}
      \pp \gets \alg{CoCo.Generate}(\secpar) \\
      \pkca \gets \alg{CA.Generate}(\secpar) \\
      \sigma \gets \text{Speranza registration for $\pkg, \id$} \\
      \cert, \pi \gets \text{Speranza signing for $\pkg, \id$} \\
      \text{s.t. } \alg{Speranza.Verify}(\pkca, \pp, \pkg, \sigma, \cert, \pi) = \yes
    \end{array}
    \right]
    = 1.
\]

\paragraph{Authenticity}
For all $\pkca, \id, \pkg$, there does not exist PPT adversary $\adv$ that can win the following game with probability greater than $\negl(\secpar)$: 
\begin{enumerate}
    \item $\pp \gets \alg{CoCo.Generate}(\secpar)$
    \item Perform package registration for $\pkg$ and $\id$
    \item $\adv$ receives $\pp$ and oracle access to $\alg{OIDC.Issue}(\skoidc, \id')$ for $\id' \neq \id$, and $\alg{CA.Issue}(\skca, \pkoidc, \cdot)$.
    \item $\adv$ outputs $\sigma, \cert, \pi$.
    \item $\adv$ wins if 
    $\alg{Speranza.Verify}(\pkca, \pp, \pkg, \sigma, \cert, \pi) = \yes$.
\end{enumerate}

\paragraph{Privacy (selective linkability)}
Let $G_0$ and $G_1$ be disjoint graph structures. For all $G_0, G_1$, for all $\id_0, \id_1$ such that $\id_0 \neq \id_1$, and for all PPT adversaries $\adv$:
\[
    \Pr \left[
    \begin{array}{l}
    \ppp \gets \alg{CoCo.Generate}(\secpar) \\
    \ppnizk \gets \alg{CoCo.GenerateEq}(\secpar) \\
    H_0 \gets \alg{FillGraph}(\ppp, \ppnizk, G_0, \id_0) \\
    b \rgets \{0, 1\} \\
    H_1 \gets \alg{FillGraph}(\ppp, \ppnizk, G_1, \id_b) \\
    b' \gets \adv(\ppp, \ppnizk, H_0, H_1) \\
    b = b'
    \end{array}
    \right]
    \leq \frac{1}{2} + \negl(\lambda).
\]
Though not intuitively linked to the Speranza protocol, this definition captures the necessary privacy requirements. As maintainers use the system, they create disjoint graphs of commitments, one graph for each package. This definition states that no matter the structure of those graphs, privacy is maintained.

\paragraph{Transparency}
We get our transparency property by applying existing techniques from key transparency to the package repository setting.
Formal analysis is left to the prior work that introduced these techniques (see \cite{parakeet}).

\subsection{Meeting Security Definitions}
The following section provides some intuition as to how Speranza meets our security definitions. 
\ifarxiv
More formal arguments can be found in \cref{sec:speranza-proof-appendix}.
\fi
\paragraph{Correctness}
This follows trivially by inspection from the correctness and completeness properties of \cocommitments{}.

\paragraph{Authenticity}
In order for an adversary to break the authenticity property, they would have to produce a \ourthing{} signature for an invalid identity that still passes all of the checks in $\alg{Speranza.Verify}$. In order for all of the checks to pass, the adversary must violate \cocommitments{}' linkability property: they have linked two unequal identities in one co-commitment. Thus, authenticity must hold.

\paragraph{Selective linkability} \label{sec: privacy}
Selective linkability reduces to the privacy of identity property of \cocommitments{}. The only difference between the privacy of identity property and this selective linkability property is that selective linkability allows for two different graph structures. The graph structure does not reveal any identity information, so selective linkability holds.

\paragraph{Transparency}
\label{sec:transparency}
The authorization record is simply an authenticated dictionary. Transparency is the ability to detect tampering with this authenticated dictionary. This is exactly the security property of an authenticated dictionary - they will not verify if there has been tampering relative to the last published digest.

Clients require that monitors sign off on the authorization record digests.
The monitors will only sign a digest after verifying \emph{all} updates to the authorization record.
Further, clients can request the history of their own packages, or any packages they depend on, and monitor those specifically by storing those policies locally.
Thus, no adversary can tamper with the authorization record without detection.

\subsection{Subverting Attacker Goals in an Honest System}

Here, we discuss attacks an attacker might attempt in pursuit of the motivations described in \cref{sec:attacker_capabilities} and the ways \ourthing{} prevents these attacks. For this section, we assume all the system components are honest and uncompromised, and an outside attacker is attempting these attacks. In \cref{sec:system_compromise}, we discuss the security of the system with different compromised components.

\subsubsection{Violate the maintainer/package mapping}
An attacker may attempt to violate the maintainer/package mapping in pursuit of running malicious code on user machines or conducting rollback attacks. This looks like signing a package on behalf of an identity the attacker does not hold. The authenticity property states that there does not exist an adversary that can publish under an identity they do not hold, so it protects against this kind of attack.

\subsubsection{Violate \ourthing{} privacy guarantees}
The selective linkability security property provides protection from this sort of attack. Selective linkability states that no matter how (poly-)many identity co-commitments are generated in whatever graph structure, parties that only have access to published graphs cannot determine what identity they belong to. 

The \ourthing{} signing protocol exposes only this graph of \cocommitments{} and a signature over the package to the public. Selective linkability means that the graph tells the public nothing. Digital signatures are generated completely independently from identities, so it is impossible for them to reveal any identity information, and adding them onto the graph structure provides no help to an adversary. Thus, outside adversaries cannot violate maintainer privacy.

\subsection{Security Under System Compromise} \label{sec:system_compromise}
The following section describes how security in this system degrades if the attacker is able to compromise elements of the system itself: the maintainer's identity provider login, an identity provider, the certificate authority, or the repository.

\subsubsection{Maintainer's identity provider login/identity provider}
As described in \cref{sec:attacker_capabilities}, this attack is out of scope, and we provide no security guarantees against an attacker that is able to compromise a maintainer's identity provider login or the identity provider itself. This is equivalent to an attacker recovering the secret key in a standard cryptographic system. We recommend implementing threshold policies for publishing in order to provide safeguards against this kind of compromise.

\subsubsection{Certificate authority}
If an attacker is able to compromise a certificate authority, \ourthing{}'s privacy guarantees fall. The attacker can see requests for certificates from maintainers and thus learn maintainer's identities. They can then learn which maintainer publishes which package by watching which certificate gets published with which package.

If the attacker has not also compromised the maintainer's package repository account, then they are still prohibited from publishing: the maintainer must authenticate with the package repository in order to publish. However, if the attacker has compromised this account \emph{and} the CA, and the attacker knows the identity of the maintainer, the attacker will be able to publish as that maintainer undetected by generating certificates to that identity and then authenticating with the package repository.

Note that there is weaker security in \ourthing{} under certificate authority compromise than in Sigstore because the privacy properties of \ourthing{} have negated the transparency properties present in the Sigstore system. In the Sigstore system, certificate authority compromise can be detected through the transparency log. However, in our system, even if all issued certificates are recorded in a tamper-proof log, there is no way to tell if those are valid or not because the subjects are all opaque. Future work is needed to provide better security under certificate authority compromise.

\subsubsection{Repository}
If an attacker is able to compromise the repository, \ourthing{}'s privacy guarantees get weaker. Since the repository maintains a mapping of package to commitment key, an attacker can now "guess and check" maintainers' identities; they can try to open the commitments in the authorization record using their guess identity and the stored commitment key. However, the attacker does not learn the maintainer's identity outright.

Even in the event of a compromised repository, an attacker cannot publish on behalf of another maintainer.
In order to publish, the attacker would need a certificate from the CA with that maintainer's identity.
This would require authenticating with an identity provider on behalf of that maintainer, which is outside of the attacker's capabilities.

Repository compromise can be detected and dealt with if the attacker attempts to alter the authorization record due to the transparency property.
However, there is no way for privacy violations by compromised repositories to be detected in \ourthing{} as long as the repository continues behaving as normal.
Further work is needed to address this problem.

\section{Implementation and Evaluation}
\label{sec:Implementation_eval}
We implemented signing and verifying for the software signing application\footnotemark.
The implementation was done in Rust version 1.63.0.
SHA2-512 was used as the hash function for all instances of hashes in the system.
For implementing Pedersen commitments, we used the Ristretto group~\cite{ristretto} over Curve25519~\cite{curve25519}.
Ristretto is a variant of Decaf \cite{decaf} that allows for a prime-order group from an elliptic curve group that actually has cofactors up to 8.
The implementation totals to \num{3700} lines of Rust code, including benchmarking harnesses.

\footnotetext{\implementationurl}

The maintainer map was implemented with a Merkle binary prefix tree, similar to the implementation described in CONIKS~\cite{coniks}.
This is consistent with the trust model and the maintainer map described in \cref{sec:simple-merkle-and-auditing}. 
All measurements were performed on a desktop computer running a \SI{3.9}{\giga\hertz} AMD Ryzen 5 5600G processor with \SI{32}{\gibi\byte} RAM running Linux.
We report the median over at least 10 trials unless otherwise mentioned.


\subsection{Signing and Verification Costs}

\begin{table}[H]
\begin{tabular}{@{}lrr@{}}
    \toprule
    \textbf{Operation} & \textbf{Create (\si{\micro\second})} & \textbf{Verify (\si{\micro\second}}) \\
    \midrule
    Ed25519 signature & \num{15} & \num{40} \\
    Pedersen commitment & \num{82} & \num{82} \\
    Pedersen equality proof & \num{188} & \num{272} \\
    Co-commitment & \num{307} & \num{362} \\
    \bottomrule
\end{tabular}
\caption{\label{tab:microbench}Microbenchmarks for cryptographic operations.}
\end{table}





\begin{figure}
    \centering
    \includegraphics[width=\linewidth]{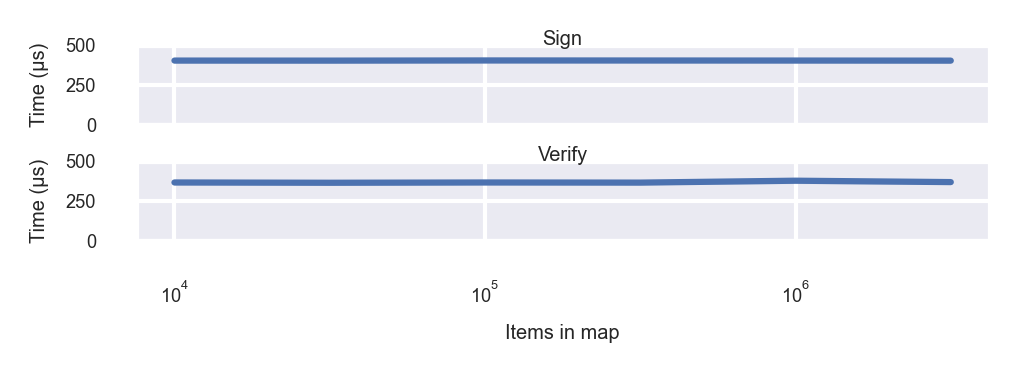}
    \caption{\label{fig:e2e-graph}%
        End-to-end costs for maintainers during package publishing (sign) and users during package download (verify).
    }
\end{figure}

We find that, for a maintainer to create a signature using identity co-commitments for a repository with 3.2 million packages requires \SI{404}{\micro\second}; verification requires \SI{372}{\micro\second}.
These numbers are consistent for repositories of varying sizes: the primary cost that changes with repository size is verifying the Merkle proof for looking up the authorization; this is small relative to the cost of verifying \cocommitments{}, and scales logarithmically with the size of the repository.
We measured, for repositories of varying size, the end-to-end signing and verification costs for maintainers and users when using the Merkle BPT-based authorization record (\cref{fig:e2e-graph}).

We report timing for the cryptographic primitives and identity co-commitments used in \ourthing{} in \cref{tab:microbench}.

\subsection{Repository Server Costs}

\begin{figure}
    \centering
    \includegraphics[width=\linewidth]{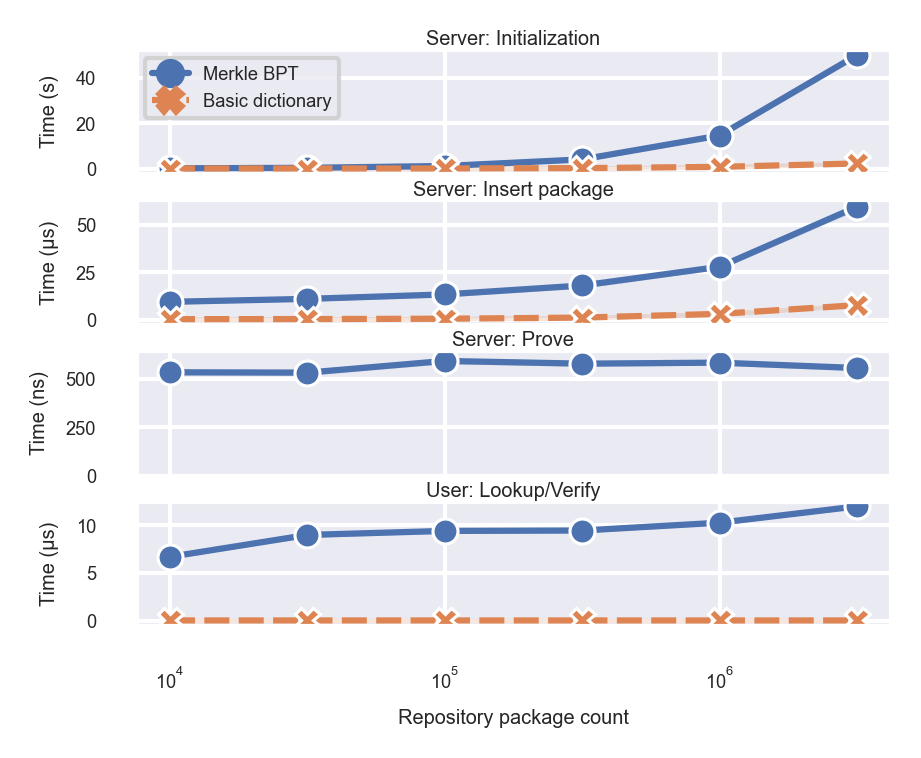}
    \caption{\label{fig:map-graph}%
        Cost of maintaining the authorization record, using both a Merkle BPT and sending the full authorization records to clients (``basic dictionary''). 
    }
\end{figure}

The primary costs to the repository server lie in maintaining the full authorization record.
In \cref{sec:simple-merkle-and-auditing}, we described two methods: first, a simple method in which users must download the complete authorization record (``basic dictionary'') and a method backed by a Merkle binary prefix trie (BPT).
We report the costs for varying operations with repositories with up to 3.2 million packages in \cref{fig:map-graph}.
First, we report the cost of initializing the authorization record.
This is a one-time cost that must be done when importing existing maintainership records, regardless of the use of \cocommitments{}.
Second, we have the cost to register a new package, or to update an existing package (``insert package'').
Third, we have the cost to create a proof that a given policy is accurate.
This can also be done for each package ahead of time, so that these proofs can be served via a CDN or mirror.
This does not apply to the basic authorization record.
Finally, we have the cost to users to verify a policy.
For the basic method, this means looking up the policy in the dictionary they already store; for the Merkle BPT method, this means verifying the Merkle proof.

\begin{figure}
    \centering
    \includegraphics[width=\linewidth]{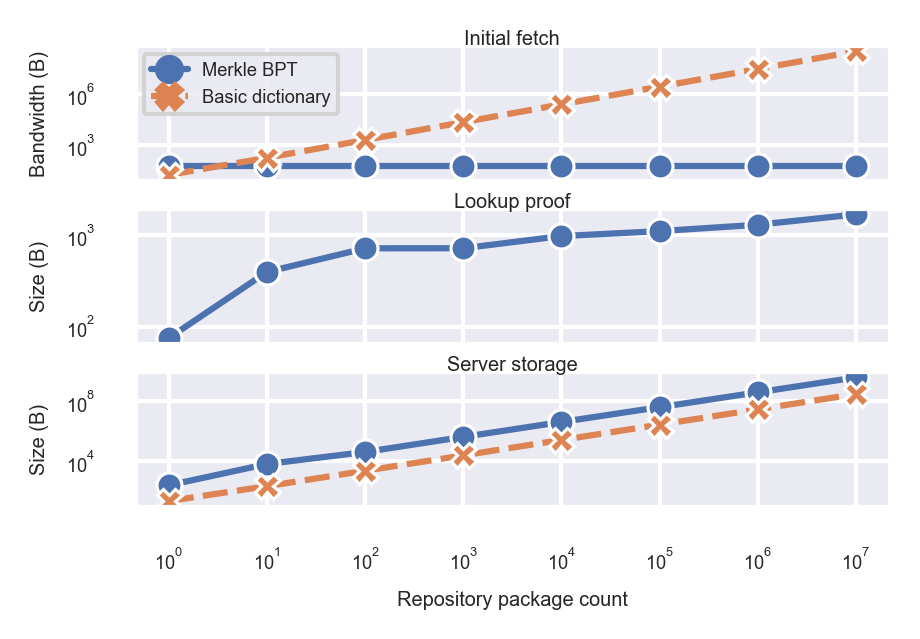}
    \caption{\label{fig:bandwidth-graph}%
    Bandwidth and storage costs in Speranza.
    }
\end{figure}

The other primary cost---and the main motivation for the more-complex Merkle BPT method---is the bandwidth used.
We report the bandwidth required and storage costs in \cref{fig:bandwidth-graph}.
The bandwidth for an initial download for the Merkle BPT method is \SI{64}{\byte}---the size of a Merkle root, which does not vary with repository size.
For the basic dictionary method, this is the size of the complete authorization record---which can be \SI{300}{\mebi\byte} for a repository of 10 million packages (assuming about \SI{32}{\byte} per entry---in practice, could be double or triple the size).
The Merkle BPT method does require ``lookup proofs'' each time a user verifies a signature, to check the policy against the authorization record.
These top out around \SI{1.6}{\kibi\byte} per package for a repository of 10 million packages.
Finally, we note the required storage costs for the repository server to maintain these records.
To maintain the Merkle BPT requires store the entire authorization record in the clear, plus some additional Merkle BPT data (sublinear in the size of the record).

\subsection{Proof of Concept: Identity Co-Commitments} \label{sec:deployability}
While our benchmarks cover the operations for signers, verifiers, and the repository server, we wanted to understand the difficulty of modifying a certificate authority to support \cocommitments{}.
We added these \cocommitments{} to the Fulcio certificate authority in the Sigstore~\cite{newman2022sigstore} project.
Our patch\footnotemark{} added one dependency, on an implement of Ristretto, and implemented Pedersen commitments and proofs of equality in \num{115} lines of Go code; to replace email addresses with commitments then required changing \num{5} lines of code. 
While a production implementation would require more testing, the changes are overall modest.
\footnotetext{\prurl}

\section{Discussion and Future Work} \label{sec:discussion_future_work}

Our evaluation shows that the costs to repositories, signers, and maintainers are reasonable for repositories of practical sizes (millions of packages and maintainers).

Empirical data about the scale of package repositories allows us to predict how our system will perform in a real-world setting.
\Cref{tab:repo_scale_table} shows the total number of packages for popular package repositories, fetched from the indicated sources on May 4, 2023.
In our evaluation, all experiments used a repository size exceeding that of npm, the largest reported repository.
We found that, aside from a one-time setup cost (under one minute), the cryptographic server operations were all sub-millisecond---less than a database write would take, and therefore negligible compared to the costs repositories already require in upkeep.
The bandwidth requirements for signatures were about \SI{1.6}{\kibi\byte}---far smaller than a typical software package.
The cost to a maintainer to sign a package is sub-millisecond---far less than the network roundtrip to the repository to publish a package.
Verification times are similar.
Even for a project depending on \emph{every} package in a repository like PyPI, verification would take only a couple of minutes, much quicker than downloading these dependencies.

\begin{table}[h]
\begin{tabular}{@{}lrr@{}}
    \toprule
    \textbf{Repository} & \textbf{Packages} & \textbf{Source} \\
    \midrule
    npm & ``2 million'' & \href{https://npmjs.org/}{npmjs.org} \\
    PyPI & \num{451913} & \href{https://pypi.org/}{pypi.org} \\
    RubyGems & \num{176365} & \href{https://rubygems.org/stats}{rubygems.org} \\
    Arch User Repository & \num{84624} & \href{https://aur.archlinux.org/}{aur.archlinux.org} \\
    Ubuntu 23.04 & \num{35587} & \href{https://repology.org/repositories/statistics}{repology.org} \\
    Hackage & \num{15903} & \href{https://hackage.haskell.org/packages/browse}{hackage.haskell.org} \\
    Arch Linux (official) &\num{13889} & \href{https://archlinux.org/packages/}{archlinux.org} \\
    \bottomrule
\end{tabular}
    \caption{Scale of real-world repositories}
    \label{tab:repo_scale_table}
\end{table}

\label{sec:pypi-changes}
We obtained data on package registrations and all changes to package ownership from PyPI.
The PyPI security team requests that we not publish this data, but encourages interested researchers to reach out to \texttt{security@pypi.org}.
The rate of changes in ownership affects operations related to maintaining the authorization record, and this data can also help predict future repository growth.
2022 saw about \num{112000} total changes to the ownership of packages.
Of those, about \num{103000} corresponded to new packages registered, leaving about \num{9000} changes to the ownership of existing packages.
This scale (12 role changes per hour) opens the door to more-expensive authenticated data structures backing the authorization record to support client verifiability (see \cref{sec:client-verifiability}).

\subsection{Related Work}
\label{sec:related-work}
Here, we discuss work on software signing and key management.

\paragraph{Packaging signing with long-lived keys.}
In traditional package signing systems, maintainers use PGP/GPG~\cite{pgp,gpg} or similar keys to sign software packages.
These systems do respect maintainer privacy, as the public keys are opaque and not linked to their identity.
However, key management can be challenging for maintainers \cite{digsigs_github, whitten99, johnny-still-cant-encrypt}, and usability concessions like account recovery in the case of lost keys can reintroduce the same security issues that signatures aim to solve.

\paragraph{Public key infrastructure for software signing.}
Using public key infrastructure (PKI) for software signing improves usability for both signers and verifiers.
Verifiers no longer need to manage a public key for each signer.
Instead, they just need to maintain a smaller root-of-trust, which they can use to verify abilities.
A certificate authority (CA) issues certificates to signers after a registration authority (RA) checks their identity.

Authenticode~\cite{authenticode} is a PKI system for software signing used on Microsoft Windows.
Though many commercial software vendors use Authenticode to sign their software,
code signing certificates for Authenticode come from a small, trusted list of CAs.
These CAs typically charge for certificates, limiting their use among hobbyists and open source maintainers.

Sigstore~\cite{newman2022sigstore} is a PKI system for software signing featuring an automated CA,
inspired by the ACME protocol~\cite{acme} used by Let's Encrypt~\cite{lets_encrypt}.
In Sigstore, certificates are linked to identities like a user account with an OpenID Connect~\cite{openid_2022} identity provider.
Sigstore can also issue certificates to machine identities: for instance, a particular build job on a continuous integration/continuous deployment (CI/CD) system.
These certificates are free and easy to obtain.
Further, because issuance is automated, signers can use a new key pair for each artifact they sign, completely obviating the need for key management.

While these PKI systems handle identity, they do not prescribe which identities should be trusted for particular software artifacts (nor, for that matter, do digital signature schemes without PKI); that requires a package repository policy.

\paragraph{Privacy-friendly credentials}
In Verifiable Credentials~\cite{verifiable-credentials}, an issuer issues credentials to holders, who \emph{present} them to verifiers.
Instead of showing the credentials directly to a verifier, a presentation can use zero-knowledge proofs of some predicate on the credentials to preserve the privacy of holders.
However, the verifier must still know \emph{what} predicate they're verifying: holders would still need to reveal their identity to prove authorization without something like \cocommitments{}.

OACerts~\cite{oacerts} use Pedersen commitments inside of certificates, and allow holders to prove predicates over their contents.
OACerts primarily supports predicates over numerical values, including support equality checks.
However, there's no notion of co-commitments---the predicates are assumed to be publicly known.

\paragraph{Package repository policy.}
A package repository policy specifies, for each package, who must sign a particular software artifact.
Centralized package repositories, like the Debian or Red Hat repositories, have a small number of trusted keys used to sign all of their packages.
A repository may want to support revocation in case of compromise, support delegation for specific packages to specific maintainers, and address a number of subtle attacks.
Naive solutions fall into a trap where verifiers download software from a source, then immediately ask the same source how to verify the software.
The Update Framework (TUF)~\cite{tuf,mercury,diplomat} handles these issues, with different roles in the system with different privileges.

TUF presumes a binary model of validation: an artifact is ``good'' or ``bad.''
The in-toto project~\cite{in-toto} allows more nuanced policies, where a package might need a signature from a a maintainer over its source, \emph{and} a signature from a trusted build service tying the ultimate artifact back to that source.
Both TUF and in-toto use long-lived keys for package repository security, though enhancement proposals to both~\cite{in-toto-sigstore-ite,tuf-sigstore-tap} propose integrating Sigstore.

CHAINIAC~\cite{chainiac} implements a software update policy with an emphasis on build reproducibility and with support for updates to authorization keys.
Much of the design of CHAINIAC aims for decentralization, a non-issue for existing package repositories, leading to higher costs, and requires user-managed keys.

\paragraph{Pseudonyms}
If maintainers are concerned about privacy, can we identify maintainers by pseudonyms?
A privacy-sensitive maintainer could register a new account for their packaging activities.
In addition to the extra hassle, however, it represents a security risk: each new account must be managed, leading to security shortcuts (like disabling multi-factor authentication, or reused passwords).
Further, this approach does not help in cases (such as harassment) where a maintainer \emph{later} decides they want to obscure their identity.

Parties in this system could provide automated pseudonyms.
The OpenID Connect protocol supports pairwise pseudonymous identifiers (PPIDs)~\cite{ppid}.
An identity provider, rather than issuing tokens with the account of a user as the subject, instead picks a distinct pseudonym for each \emph{audience}.
For example, the provider could use $\hash(\mathsf{sk},  \mathsf{user}, \mathsf{aud})$, where $\hash(\cdot)$ is a collision-resistant hash function, $\mathsf{user}$ is a username, like $\mathtt{user@example.com}$, $\mathsf{aud}$ is the audience, such as $\mathtt{sigstore.dev}$, and $\mathsf{sk}$ is a secret, provider-wide salt (preventing brute-force guessing of the user name).
Automated pseudonyms require support from upstream identity providers; few major providers implement them.

If the identity provider does not support PPIDs, the certificate authority could introduce automated pseudonyms themselves.
Creating pseudonyms from a verifiable random function (VRF)~\cite{vrf}, as in CONIKS~\cite{coniks}, allows someone knowing the cleartext identity, a shared ``verifying key'' for the CA, and a non-interactive proof to verify pseudonym correctness.
However, using a VRF in this manner requires revealing the cleartext identity to any user wishing to verify it (in this setting, just the package repository); \cocommitments{} are verifiable \emph{without} knowing the identity.
As in \ourthing{}, this limits the ability to monitor the certificate authority for specific identities, though the VRF holder \emph{can} learn their identity's VRF output and scan for it.
Further, the fact that a VRF is deterministic for a given key means that a given identity has only one pseudonym for the given ecosystem, allowing correlation across services; multiple VRFs could be used at the cost of extra complexity.
A given ecosystem would be tied to a single VRF key, requiring the user to sign with the same CA each time.

\paragraph{Transparency systems}
In a transparency system, a centralized party maintains a tamper-proof, public log.
These systems are appropriate in ``trust-but-verify'' settings: unlike in a blockchain, there is a centralized party that can modify the log at will.
This centralization avoids the need for expensive consensus mechanisms.
Despite centralization, these systems are \emph{accountable}: if the centralized party misbehaves, it can be detected and publicized.
These systems use cryptographic techniques to ensure that no data is ever removed from the log and verify.

The transparency system with widespread deployment was certificate transparency~\cite{ct}, which logs all certificates issued in web PKI.
Several projects add transparency for software (Go's sumdb~\cite{sumdb}, Firefox's binary transparency~\cite{firefox-bin-trans}) or signatures (Sigsum~\cite{sigsum}).
Sigstore~\cite{newman2022sigstore} features transparency logs for both certificates issued and signatures published.


\paragraph{Key transparency}
Several recent works aim to provide a map from identities to public keys (often in support of end-to-end user chat applications), using transparency techniques for accountability.
CONIKS~\cite{coniks} uses a verifiable random function~\cite{vrf} to anonymize user identities along with a Merkle prefix tree to map identities to keys.
Third-party monitors verify the \emph{consistency} of the data structure and prevent equivocation, but do not check for the correctness of changes.
Instead, each user monitors their own key history.
The VRF used for privacy in CONIKS inspired \cocommitments{}.

Since CONIKS, a number of key transparency systems with interesting properties have emerged.
Gossamer~\cite{gossamer} manages package ownership on a transparency log, requiring auditors and clients to walk the entire log for updates.
Google's Key Transparency~\cite{googel-key-transparency}, since abandoned, is an implementation of key transparency closely following CONIKS.
SEEMless~\cite{seemless} improves the performance and scaling characteristics of CONIKS.
Verdict~\cite{verdict} proposes a new implementation of a transparency dictionary, which can be used to construct a key transparency system.
Parakeet~\cite{parakeet} improves on the consistency mechanism required for key transparency and achieves sufficient performance to support billions of users; an implementation powers key transparency for WhatsApp~\cite{parakeet-whatsapp}.


While the techniques from key transparency systems are useful for package ownership management, there are a few key differences.
First, the scale of key transparency systems (Parakeet handles billions of users) is far greater than required for package repository.
Second, the correctness of updates to a package's authorization policy can be verified by third parties, while key updates cannot (to allow for lost keys).
Third, the privacy considerations are different: package names and package authorization policies are both are public (it is just the identities which are hidden) which enables any client to audit any package.
Together, this allows \ourthing{} auditors to verify the correctness of every update to every package, which would be prohibitively expensive for key transparency. 

Further, while key transparency systems do support mechanisms for privacy, these mechanisms apply to the \emph{labels} of the directory (usernames or phone numbers). 
In the package repository setting, these \emph{labels} are public, but the cryptographic identities themselves must be hidden.
The techniques for privacy in the key transparency setting do not apply out-of-the box to private signing with certificates, which motivates our \cocommitments{}.

\subsection{Auditing the Authorization Record}
\label{sec:discussion-transparency}
Signatures, even those checked against an authorization record are only useful if that authorization record is correct:
a compromised repository could serve a bad authorization record to a user.

Key transparency systems~\cite{coniks,seemless,parakeet} have identified three potential mitigations for such attacks on a different type of directory, which maps identities to public keys.
First, a \emph{consistency protocol} ensures that the server cannot equivocate and serve different views of the  directory to different clients: clients require signatures from a quorum of distributed witnesses.

Second, updates to the record should be publicly auditable, so that third party \emph{monitors} will notice any attempts to tamper with the history of the record.
Third, identity owners should be able to efficiently check their own keys for unauthorized modifications (\emph{correctness}).

\ourthing{} achieves transparency through similar mechanisms, though with slight modifications.
The consistency protocol applies exactly as before, and we recommend the use of the protocol used in Parakeet~\cite{parakeet}.
However, in key transparency systems, the monitors perform only checks of \emph{consistency}, not of the \emph{correctness} of updates.
Key transparency systems manage billions of keys, and checking each individual update is infeasible.
Except in CONIKS's optional ``paranoid'' mode \cite{coniks} which disables replacing lost keys, correctness can \emph{only} be verified by end-users: replacing a lost key looks like unauthorized tampering.
In \ourthing{}, it is practical for monitors to check each update (about 12 per hour; see \cref{sec:pypi-changes}).
Individuals \emph{can} audit the history for packages, but since package maintainers typically only interact with a repository when they publish a release, global correctness monitoring avoids the requirement for each package maintainer to periodically come online.
Further, because the \emph{labels} (package names) are cleartext, verifiers can check correctness for packages they would like to use by requesting the history from the repository and verifying the history's consistency.

\paragraph{Client verifiability}
\label{sec:client-verifiability}
In \ourthing{}, monitors audit both the correctness of updates to the authorization map and the consistency of the map itself.
That is, the package repository should not be able to tamper with the history of any package, nor should it be able to make changes to a packages ownership inconsistent with its current policy.
In the version of \ourthing{} with complete authorization records, clients rely on monitors for consistency.
Maintainers and verifiers can, if desired, check the correctness of individual packages they care about by storing the current state for those packages locally.

We note that at the scale of package repositories (using data from PyPI as a test case), \emph{global} correctness checking is possible.
First, if clients download the full authorization map, they can request a list of updates to the map and verify each one (at a rate of only about \num{100000} per year).
To save on bandwidth for the initial download, along with storage costs for the client, we could use \emph{transparency dictionaries}~\cite{verdict,versa}, which are exactly the required data structure: each package has its own entry, and the \emph{history} of updates to each package are append-only.
Unfortunately, current transparency dictionaries have somewhat higher latency, especially for updating, than would be appropriate for a production deployment.
We note the setting where relatively few \emph{updates} to package ownership occur relative to the total number of packages admits an interesting optimization, which we explore in 
\ifarxiv
\cref{sec:rsa-record-appendix}.
\fi

\subsection{Future Work} \label{sec:future-work}
We note the following directions for future work.
As discussed in \cref{sec:system_compromise}, compromise of either of the certificate authority or repository can  cause \ourthing{}'s privacy guarantees to fail.
Furthermore, detection of this compromise is difficult, and the certificate authority can issue malicious certificates without detection under \ourthing{}, because identities are no longer cleartext.
More work in this space is needed to protect against this kind of event.
Redaction mechanisms for transparency logs may help: if information is public by default, but redactable on request, it is transparent (but not private) in real-time but supports privacy later.
Further zero-knowledge proofs could support monitoring use cases.

We noted in \cref{sec:client-verifiability} that clients could individually audit for the correctness of updates to the policy for a particular package and propose an example architecture supporting this in 
\ifarxiv
\cref{sec:rsa-record-appendix}.
\fi
Future work might search more for specific cryptographic primitives to construct this architecture, and fully evaluate a system supporting global correctness auditing by clients.

\section{Conclusion}
\label{sec:conclusion}
We present \ourthing{}, a system for usable (certificate-based) anonymous software signing. We present \cocommitments{}, a technique for one-time-use pseudonyms using Pedersen commitments and Chaum-Pedersen proofs of commitment equality. We then use \cocommitments{} and data structure techniques from key transparency to construct our anonymous software signing system. The system is fast, practical, and easily deployable. Future work is needed for better compromise detection and client verifiability of the repository.

\begin{acks}
This research was partially supported by the NSF under Award No.\ 2229703 and Cisco Research.
We thank the anonymous reviewers for many helpful suggestions, and Nikos Vasilakis for insightful discussions.
\end{acks}

\bibliographystyle{ACM-Reference-Format}
\bibliography{main.bib}

\appendix

\ifarxiv
\section{Basic Cryptographic Definitions and Constructions} \label{sec:crypto_appendix}
Here, we give definitions and constructions for cryptographic commitments, Pedersen commitments, proofs of commitment equality, and digital signatures.

\subsection{Commitment Schemes}

Commitments provide a cryptographic "lock box", allowing one party to put information inside the box, give the box to a second party, and then later give the second party the key to "open" the commitment.

A commitment scheme as the following three algorithms:
\begin{itemize}
  \item $\alg{Com.Generate}(\secpar) \to \pp$: creates commitment scheme public parameters.
  \item $\alg{Com.Commit}(\pp, m) -> (c, r)$: generates a commitment to value $m$ and returns the commitment as well as the random key $r$.
  \item $\alg{Com.Verify}(\pp, m, c, r) -> \yes / \no$: verifies the opening of commitment $c$ with message $m$ and key $r$.
\end{itemize}

\noindent
The properties of commitment scheme are as follows:
\begin{itemize}
    \item \textit{Correctness.}
    For all $m$:
    \[
        \Pr \left[ 
        \begin{array}{l}
            \pp \gets \alg{Com.Generate}(\secpar) \\
            (c, r) \gets \alg{Com.Commit}(\pp, m) \\
            \alg{Com.Verify}(\pp, m, c, r) = yes
        \end{array}
        \right]
        = 1.
    \]

    \item \textit{Hiding.}
    For all $m_0, m_1$ and for all $\ell$ that are polynomial in $\secpar$:
    \[
        \Pr \left[ 
        \begin{array}{l}
            \pp \gets \alg{Com.Generate}(\secpar) \\
            b \rgets \{0, 1\} \\
            (c_i, r_i) \gets \alg{Com.Commit}(\pp, m_b) \\
            \adv(\secpar, \pp, c_0 ... c_{\ell}) = b \\
        \end{array}
        \right]
        \leq \negl(\secpar).
    \]
    
    \textit{Note: the traditional definition of hiding only includes one sampled commitment instead of poly-many. However, since $\alg{Pedersen.Commit}$ can be computed in poly-time, these are equivalent.} 
    
    \item \textit{Binding.}
    \[
        \Pr \left[ 
        \begin{array}{l}
            \pp \gets \alg{Com.Generate}(\secpar) \\
            (c, m_1, r_1, m_2, r_2) \gets \adv(\secpar, \pp) \\
            m_1 \neq m_2 \text { AND } \\
            \alg{Com.Verify}(\pp, m_1, c, r_1) = \yes \text { AND } \\ 
            \alg{Com.Verify}(\pp, m_2, c, r_2) = \yes
        \end{array}
        \right]
        \leq \negl(\secpar).
    \]
\end{itemize}

\subsubsection{Pedersen commitments}
Pedersen commitments \cite{Pedersen_commits} are a commitment scheme based on the discrete log assumption. 

Assume there exists a group $\mathbb{G}$ of order $q$ where $q$ is a large prime. Let $g$ and $h$ be two generators of $\mathbb{G}$. $g$ and $h$ must be chosen in a trusted setup phase such that $\log_g h$ is unknown. Under the discrete log assumption, since $\mathbb{G}$ is of prime order, $g$ and $h$ can be chosen randomly. The choice of $\mathbb{G}$, $g$, and $h$ should be done in a trusted setup phase.

To commit to $x \in \mathbb{Z}_q$, first draw randomness $r \leftarrow \mathbb{Z}_q$. Then, compute
\begin{displaymath}
    c = g^x h^r.
\end{displaymath}

Since $r$ is random, $h^r$ acts essentially like a one-time pad, hiding $g^x$. The scheme is computationally binding as if an adversary can compute $log_g h$, they can open a commitment to whatever $x'$ they like. However, by the discrete log assumption, no computationally bounded adversary can compute this, so the scheme is binding. Formal proofs are left to Pedersen's paper \cite{Pedersen_commits}.

\subsection{Zero-Knowledge Proof-of-Knowledge Proof of Commitment Equality}
A proof of commitment equality is a zero-knowledge proof-of-knowledge convincing a verifier that (a) these two commitments are commitments to the same message and (b) the prover knows what that message is.

Given an efficiently-computable proof of commitment equality, a commitment scheme can be augmented with three more algorithms:
\begin{itemize}
  \item $\alg{Com.GenerateEq}(\secpar) \to \ppnizk$.
  \item $\alg{Com.ProveEq}(\pp, \ppnizk, m, c_1, r_1, c_2, r_2) \to \pi \text{ or } \bot$.
  \item $\alg{Com.VerifyEq}(\pp, \ppnizk, c_1, c_2, \pi) \to \yes / \no$.
\end{itemize}

\noindent
These proofs have the following security properties:
\begin{itemize}
    \item \textit{Non-Interactive Zero Knowledge (NIZK).}
    There exists poly-time algorithms $\simu_1, \simu_2$ such that 
    for all $\pp$, $c_1, c_2, m, r_1, r_2$ satisfying
    \[c_1, r_1 = \alg{Com.Commit}(\pp, m)\] and \[c_2, r_2 = \alg{Com.Commit}(\pp, m),\]
    and for all PPT adversary $\adv$,
    $\adv$ wins the following game ($b = b'$) with probability $\leq \frac{1}{2} + \negl(\secpar)$:
    
    \begin{enumerate}
        \item Challenger draws $b \rgets \{0,1\}$
        \item If $b = 0$:
        \begin{enumerate}
            \item $\ppnizk \gets \alg{Com.GenerateEq}(\lambda)$
            \item $\pi \gets \alg{Com.ProveEq}(\pp, \ppnizk, m, c_1, r_1, c_2, r_2)$
        \end{enumerate}
        \item If $b = 1$:
        \begin{enumerate}
            \item $(\ppnizk, \aux) \gets \simu_1(\secpar)$
            \item $\pi \gets \simu_2(\pp, \ppnizk, \aux, c_1, c_2)$
        \end{enumerate}
        \item $b' \gets \adv(\ppnizk, \pi)$
    \end{enumerate}
  
    \item \textit{Completeness.}
    For all $\pp$, $c_1, c_2, m, r_1, r_2$,
    satisfying \[c_1, r_1 = \alg{Com.Commit}(\pp, m)\]
    and \[c_2, r_2 = \alg{Com.Commit}(\pp, m),\]
    we have:
    \[
        \Pr \left[ 
        \begin{array}{l}
            \ppnizk \gets \alg{Com.GenerateEq}(\secpar) \\
            \pi \gets \alg{Com.ProveEq}(\pp, \ppnizk, m, c_1, r_1, c_2, r_2) \\
            \alg{Com.VerifyEq}(\pp, \ppnizk, c_1, c_2, \pi) = \yes
        \end{array}
        \right]
        = 1.
    \]
    
    \item \textit{Knowledge Soundness.}
    For all $m, r_1, r_2, c_1, c_2$, there exists a PPT algorithm $\ext$ such that, given oracle access to $\hat{P}$ \\
    (which may be $\alg{Com.ProveEq}(\cdot, \cdot, \cdot, \cdot, \cdot, \cdot, \cdot)$): \\
    \[
        \Pr \left[ 
        \begin{array}{l}
            \pp \gets \alg{Com.Generate}(\secpar) \\
            \ext(\pp) \to (m', r_1', r_2') \\
            \alg{Com.Verify}(\pp, m', c_1, r_1') = \yes \text{ AND } \\
            \alg{Com.Verify}(\pp, m', c_2, r_2') = \yes \\
        \end{array}
        \right]
        - \\ \]
        \[
        \Pr  \left[ 
        \begin{array}{l}
            \pp \gets \alg{Com.Generate}(\secpar) \\
            \ppnizk \gets \alg{Com.GenerateEq}(\secpar) \\
            \pi \gets \alg{Com.ProveEq}(\pp, \ppnizk, m, c_1, r_1, c_2, r_2) \\
            \alg{Com.VerifyEq}(\pp, \ppnizk, c_1, c_2, \pi) = \yes \\
        \end{array}
        \right]
        \leq \negl(\secpar).
    \]  
\end{itemize}

\subsubsection{Pedersen commitment equality}
Zero-knowledge proof-of-knowledge proof of equality for pedersen commitments can be done using the Chaum-Pedersen protocol \cite{ChaumPedersenProof}. The protocol requires that the prover know the message committed to by both commitments as well as the random keys for both commitments. The protocol goes as follows.

Let
\begin{align*}
    c_1 &= g^x h^{r_1} \\
    c_2 &= g^x h^{r_2}
\end{align*}
where $c_1$ and $c_2$ are public, and $x, r_1, \text{ and } r_2$ are private. The prover and verifier then exchange the following messages:

\begin{center}
    \begin{tabular}{c|c c c}
         & \textbf{P} & & \textbf{V}  \\ \hline
        1 & $s_1, s_2, s_3 \leftarrow \Z_q$ & & \\
         & $\alpha_1 = g^{s_1}h^{s_2}, \alpha_2 = g^{s_1}h^{s_3}$ & & \\
        2 & & $\xrightarrow{\alpha_1, \alpha_2}$ & \\
        3 & & & $d \leftarrow \Z_q$ \\
        4 & & $\xleftarrow{d}$ & \\
        5 & $\beta_1 = dx + s_1$ & & \\
         & $\beta_2 = dr_1 + s_2$& & \\
         & $\beta_3 = dr_2 + s_3$ & & \\
        6 & & $\xrightarrow{\beta_1, \beta_2, \beta_3}$ & \\
        7 & & & $\alpha_1 \cdot c_1^d \stackrel{?}{=} g^{\beta_1} \cdot h^{\beta_2}$ \\
         & & & $\alpha_2 \cdot c_2^d \stackrel{?}{=} g^{\beta_1} \cdot h^{\beta_3}$ \\
    \end{tabular}
\end{center}

This protocol has soundness error $\frac{1}{q}$. As long as $q$ is sufficiently large, only one round of this protocol is necessary to achieve negligible soundness error.

The proof can then be made non-interactive using the Fiat-Shamir heuristic \cite{fiat_shamir}. Instead of getting the challenge from the verifier in step 3, the prover instead computes $d = H(c_1, c_2, \alpha_1, \alpha_2)$ where $H$ is a collision-resistant hash function to $\Z_q$. The verifier then checks that $d$ was computed correctly when checking the proof. 
In the random oracle model, $d$ is now just as random and unpredictable as if the verifier had drawn it randomly from $\Z_q$.
Since the hash is collision-resistant, the prover cannot find another input to generate the same challenge and thus cannot ``work backwards.''
In this model, the hash function $H$ is the "public parameter" $\ppnizk$.

\subsection{Digital Signatures}
We define the following API for digital signatures:
\begin{itemize}
    \item $\alg{DigSig.Generate}(\secpar) \to (\sk, \pk)$
    \item $\alg{DigSig.Sign}(\sk, m) \to \sigma$
    \item $\alg{DigSig.Verify}(\pk, m, \sigma) \to \yes / \no$
\end{itemize}

\noindent
Digital signatures then have the following security properties:
\begin{itemize}
    \item \textit{Correctness.} For all $m$:
    \[
    \Pr \left[ 
      \begin{array}{l}
        (\sk, \pk) \gets \alg{DigSig.Generate}(\secpar) \\
        \sigma \gets \alg{DigSig.Sign}(\sk, m) \\
        \alg{DigSig.Verify}(\pk, m, \sigma) = \yes
      \end{array}
    \right]
    = 1
    \]

    \item \textit{Unforgeability.}
    For all PPT adversaries $\adv$, and for all $\ell$ such that $\ell$ is polynomial in $\secpar$, $\adv$ wins the following game with probability $\leq \negl(\secpar)$:

    \begin{enumerate}
        \item Challenger runs $(\sk, \pk) \gets \alg{DigSig.Generate}(\secpar)$ and sends $\pk$ to $\adv$
        \item Repeat $\ell$ times:
        \begin{enumerate}
            \item $\adv$ sends Challenger $m_i$
            \item Challenger computes $\sigma_i \gets \alg{DigSig.Sign}(\sk, m_i)$; sends $\sigma_i$ to $\adv$
        \end{enumerate}
        \item $\adv$ returns $(m^*, \sigma^*)$
        \item $\adv$ wins if $(m^*, \sigma^*) \notin \{ (m_i, \sigma_i) \}$
    \end{enumerate}
    
\end{itemize}

\section{Identity Co-Commitments Security Proofs} \label{sec:coco-proofs-appendix}
In this section, we give formal arguments for identity co-commitment security properties.
\subsection{Privacy of Identity}
    For all $\id_0, \id_1, G$,
    where $G$ is an undirected graph structure:
    \begin{align*}
        (\ppp, &\ppnizk, \alg{FillGraph}(\ppp, \ppnizk, G,\id_0)) \approx_c \\
        &(\ppp, \ppnizk, \alg{FillGraph}(\ppp, \ppnizk, G,\id_1))
    \end{align*}

    \noindent\textbf{Proof:}
    This proof is a hybrid argument, starting with the left side of the computational indistinguishability claim above and working toward the right side.
    
    \begin{enumerate}
    \item Start with $(\ppp, \ppnizk, G_0)$: \\
       - $\ppp \gets \alg{CoCo.Generate}(\secpar)$ \\
       - $\ppnizk \gets \alg{CoCo.GenerateEq}(\secpar)$ \\
       - $c_i, r_i \gets \alg{CoCo.Commit}(\ppp, \id_0)$ for all nodes\\
       - $e_{ij}$ from $\alg{CoCo.ProveEq}(\ppp, \ppnizk, \id_0, c_i, r_i, c_j, r_j)$ for all edges \\
       - Return $\ppp, \ppnizk, G$ \\
       This is equivalent to the left side of the computational indistinguishability claim. \\
       
    \item Recompute $G_0$ with simulated NIZK public parameters: \\
       - $\ppp \gets \alg{CoCo.Generate}(\secpar)$ \\
       - $\ppnizk, \aux \gets \simu_1(\secpar)$ \\
       - $c_i, r_i \gets \alg{CoCo.Commit}(\ppp, \id_0)$ for all nodes\\
       - $e_{ij}$ from $\alg{CoCo.ProveEq}(\ppp, \ppnizk, \id_0, c_i, r_i, c_j, r_j)$ for all edges \\
       - Return $\ppp, \ppnizk, G_0$

       If there exists a PPT adversary that can distinguish between (1) and (2), there exists an adversary to win the Pedersen NIZK game with non-negligible probability.

       Suppose towards contradiction that there exists some adversary $\adv$ that can tell (1) from (2). \\
       Let $c', r' = \alg{CoCo.Commit}(\id')$, \\
       Let $c'', r'' = \alg{CoCo.Commit}(\id')$, \\
       Then we build adversary $\advb$ that can break NIZK for \\
       $(\ppp, c', c'', \id', r', r'')$:

       On input $(\ppnizk, \pi)$, $\advb$ takes the following steps: \begin{enumerate}
           \item Discard $\pi$
           \item Build graph $G$. Compute nodes with $\alg{CoCo.Commit}$ and use  the received $\ppnizk$ to compute edges
           \item Give this graph to $\adv$. If $\ppnizk$ was honestly, generated by the challenger, $\adv$ sees exactly (1). Otherwise, $\adv$ sees exactly (2).
           \item $\adv$ distinguishes real from simulated, and $\advb$ forwards $\adv$'s response to the challenger and wins
       \end{enumerate}

    \vspace{1cm}
    
    \item (repeat this step of the hybrid for each edge $e_{ij}$) Recompute with $e_{ij}$ swapped for \emph{simulated} edge. \\
       - $\ppp, \gets \alg{CoCo.Generate}(\secpar)$ \\
       - $\ppnizk \gets \alg{CoCo.GenerateEq}(\secpar)$ \\
       - $\ppnizk, \aux \gets \simu_1(\secpar)$ \\
       - $c_i, r_i \gets \alg{CoCo.Commit}(\ppp, \id_0)$ \\
       - Edges are computed the same way as in previous step, except $e_{ij}$ is now $\simu_2(\ppnizk, \aux, c_i, c_j,)$ \\
       - Return $\ppp, \ppnizk, G$

       If there exists a PPT adversary that distinguish between step $3t$ and $3(t-1)$, there exists an adversary that can win the Pedersen NIZK game with non-negligible probability.

       Suppose towards contradiction there exists some adversary $\adv$ could distinguish.
       Now we're going to build adversary $\advb$ for NIZK with $\ppp, c_i, c_j, \id_0, r_i, r_j$:

       On input $(\ppnizk, \pi)$, $\advb$ takes the following steps:
       \begin{enumerate}
           \item Build graph $G$. Compute nodes with $\alg{CoCo.Commit}$, and use the received $\ppnizk$ OR $\simu_2$ for edges as appropriate. Use $c_i, c_j$ for nodes, and use $\pi$ instead of computing something new for $e_{ij}$
           \item Give $G$ to $\adv$. If $\pi$ was computed honestly, this is exactly step $3(t-1)$, otherwise, it is step $3t$.
           \item $\adv$ distinguishes real from simulated, and $\advb$ forwards $\adv$'s response to the challenger and wins
       \end{enumerate}
       
    \vspace{1cm}

    \item Swap out all of the commitments from $\id_0$ to $\id_1$. Simulate new proofs. \\
       - $\ppp \gets \alg{CoCo.Generate}(\secpar)$ \\
       - $\ppnizk, \aux \gets \simu_1(\secpar)$ \\
       - $c_i, r_i \gets \alg{CoCo.Commit}(\ppp, \id_1)$ for all nodes\\
       - $e_{ij} \gets \simu_2(\ppnizk, \aux, c_i, c_j)$ for all edges \\
       - Return $\ppp, \ppnizk, G$

    If there exists an adversary that can distinguish the last step of (3) from (4), there exists an adversary that can win the Pedersen hiding game with non-negligible probability.

    For $\ell = |G|$: \\
    Suppose towards contradiction there exists some adversary $\adv$ could distinguish. Now we're going to build adversary $\advb$ for hiding.

    On input $c_1, \dots, c_{\ell}$, $\advb$ takes the following steps: \\
     - Make $c_1, \dots, c_{\ell}$ the graph vertices \\
     - Compute $\ppnizk, \aux = \simu_1(\secpar)$
     - Compute edges $e_{ij} = \simu_2(\ppp, \ppnizk, \aux, c_i, c_j)$ \\
     - $\adv$ distinguishes which identity is held in the commitments, returns to $\advb$, who forwards this to the challenger and wins. \\

    \item Go backwards to $G_1$: swap out simulated proofs for real ones by running the process in step 3 backwards, then swap out simulated parameters for real ones as in step 2, and then arrive at $G_1$. Indistinguishability holds by symmetry from the arguments above.
    
    \end{enumerate}

$\blacksquare$

\subsection{Linkability}
    For all connected graph structures $G$, and all PPT adversaries $\adv$:
    \[
        \Pr\left[
        \begin{array}{l}
            \pp \gets \alg{CoCo.Generate}(\secpar) \\
            (H, \id_i, r_i) \gets \adv(\ppp) \\
            |\{ \id_i \}| > 1 \text{ AND } \\
            \forall e_{ij}, \alg{CoCo.VerifyEq}(\pp, c_i, c_j, e_{ij}) = \yes \text{ AND}\\
            \forall \id_i, c_i, r_i, \alg{CoCo.Verify}(\pp, \id_i, c_i, r_i) = \yes
        \end{array}
        \right]
        \leq \negl(\secpar).
    \]

    \noindent\textbf{Proof:}
    Suppose towards contradiction that there exists an adversary $\adv$ that wins the above game with non-negligible probability. \\
    
    Since the graph is connected, there must be at least one pair of nodes with an edge (call them $c_1, c_2, e$) in the returned graph such that $\id_1 \neq \id_2$, $\alg{CoCo.VerifyEq}(\ppp, \ppnizk, c_1, c_2, e) = \yes$, $\alg{Pedersen.Verify}(\ppp, \id_1, c_1, r_1) = \yes$, and \\
    $\alg{Pedersen.Verify}(\ppp, \id_2, c_2, r_2) = \yes$. \\
    
    Looking at the Pedersen knowledge soundness definition, this means that for $\id_1, \id_2, c_1, c_2, r_1, r_2$, the second probability is non-negligible. Thus, in order for knowledge soundness to hold, there must also exist a PPT algorithm $\ext$ that succeeds with non-negligible probability for these values. \\

    Then we can construct $\advb$ that can win the binding game. On input $\ppp$:
    \begin{enumerate}
        \item $m', r_1', r_2' \gets \ext(\ppp)$
        \item Select $\id_x$ such that $\id_x \in \{\id_1, \id_2\}$ and $m' \neq \id_x$.
        \item $\advb$ returns $c_x, m', r_x', \id_x, r_x$
    \end{enumerate}

    Since $\ext$ succeeds with non-negligible probability, then $\Pr[\alg{Pedersen.Verify}(\ppp, m', c_x, r_x') = \yes] > \negl(\secpar)$. From the returned graph, we know that $\alg{Pedersen.Verify}(\ppp, \id_x, c_x, r_x) = \yes$. Thus, $\advb$ has won the binding game with the same probability of $\ext$'s success, which must be non-negligible.

$\blacksquare$

\section{Speranza Security Proofs}
\label{sec:speranza-proof-appendix}
Here, we provide formal arguments of Speranza's security.
\subsection{Authenticity}
Suppose towards contraction that there exists an adversary $\adv$ that can win the above game with non-negligible probability. With $\adv$, we can construct $\advb$ that wins the \cocommitments{} linkability game with non-negligible probability.
    
If $\adv$ wins the above game, then all of the checks in $\alg{Speranza.Verify}$ must go through for $\adv$'s output.
\begin{enumerate}
    \item The certificate must verify for some subject $\sub$. This means that $\sub$ must be a commitment to some identity $\id'$.
    \item $\id' \neq \alg{AuthRecord.Lookup}(\pkg)$ because: \\
    - $\sub$ must be coming from one of the oracle calls to $\alg{CA.Issue}$ as $\adv$ cannot generate a valid certificate independently by the unforgeability of digital signatures.  \\
    - Let $\tok$ the token that was used in that call to $\alg{CA.Issue}$. $\tok$ must come from one of the oracle calls to $\alg{OIDC.Issue}$ as the adversary cannot generate a valid token independently, again by the unforgeability of digital signatures. \\
    - Since $\tok$ must have come from an oracle call to $\alg{OIDC.Issue}$, $\alg{OIDC.Verify}(\pkoidc, \tok) \neq \id$.
    \item Lastly, $\alg{CoCo.Verify}(\ppp, \ppnizk, \pkg, \sigma, \cert, \pi)$ must return $\yes$. This means $\adv$ has constructed a valid 2-node graph that violates linkability of Identity Cocommitments. $\advb$ can use $\adv$ to win the linkability game for graph structure $G$ that is two nodes connected with one edge. On input $\ppp, \ppnizk$: \begin{enumerate}
        \item Run setup for $\advb$'s game. Draw $(\skca, \pkca) \gets \alg{CA.Generate}(\secpar)$ and $(\skoidc, \pkoidc) \gets \alg{OIDC.Generate}$. Select $\id, \pkg$ and register with the repository. During registration, receive $r$.
        \item Send $\ppp, \ppnizk$ to $\adv$.
        \item Answer the adversary's queries using the $\alg{OIDC}$ and $\alg{CA}$ instance generated in step 1. Record all of the adversary's queries as $(\id'_i, r'_i)$.
        \item $\adv$ returns $\sigma, \cert, \pi$. Discard $\sigma$.
        \item $c_1 = \alg{AuthRecord.Lookup}(\pkg)$
        \item $c_2 = \alg{X509.VerifyCert}(\pkca, \cert)$
        \item As discussed above, one of the adversary's queries to $\alg{OIDC.Issue}$ must have been used to form $\cert$. Find this query. Check $\alg{Pedersen.Verify}(\ppp, \id'_i, c_2, r'_i)$ until some $(\id', r')$ pair returns $\yes$. Call this pair $(\id^*, r^*)$.
        \item Construct $H$ with nodes $c_1, c_2$ and edge $\pi$
        \item Return $H, \{\id, \id^*\}, \{r, r^*\}$
    \end{enumerate}
    As previously discussed, $\id \neq \id'$, the proof verifies, and the commitments must validate. $\advb$ wins the game, presenting a contradiction.
\end{enumerate}
$\blacksquare$

\subsection{Selective Linkability}
Suppose towards contradiction there exists a PPT adversary $\adv$ that can win the above game with non-negligible advantage. Then, we can construct adversary $\advb$ that can break \cocommitments{} Privacy of Identity.

On input $\ppp, \ppnizk, H$ to $\advb$:
\begin{enumerate}
    \item Let the structure of $H$ be $G_1$
    \item Pick any graph structure $G_0$ such that it is disjoint from $G_1$. Compute $H_0 = f(G_0, \id_0)$.
    \item Send $\ppp, \ppnizk, H_0, H_1$ to $\adv$. Return $\adv$ output to the Challenger and win.
\end{enumerate}
$\blacksquare$
\fi

\section{Anonymized Authorization Record Examples} \label{sec:example-appendix}
This appendix provides concrete examples of what verifying a policy or changing a policy might look like with an anonymized authorization record.

\subsection{Example: Verifying with a Threshold Policy}
Assume there exists artifact \texttt{Foo} that has three signers, Alice, Bob, and Charlie. Their policy is such that in order to publish any changes to \texttt{Foo}, at least two of the three signers must sign off on the change.

The following is an example of what $\alg{CheckPublish}$ might look like for an artifact:
\begin{verbatim}
struct ThresholdPolicy {
    signers: List<Commit>,
    threshold: int
}
\end{verbatim}

The input consists of at least \texttt{threshold} commitments to signers and two equality proofs.
To check publishing authorization, check, that the inputs validate for at least \texttt{threshold} distinct \texttt{signers}.

\subsection{Example: Changing Policies}
Again assume there exists artifact \texttt{Foo} with signers Alice, Bob, and Charlie. Their ownership change policy is that Alice is the head signer, and she and she alone must sign off on any changes in ownership. 

The following is an example of what the $\alg{CheckPolicyChange}$ function may look like for this policy:
\begin{verbatim}
struct HeadsignerPolicy {
    head_signer = Commit(Alice),
    signers = List<Commit>,
}
\end{verbatim}

The input consists of a commitment to the head signer, an equality proof, and a new policy. To check authorization, check the identity co-commitment between the input and the \texttt{head\_signer}. If it holds, overwrite the policy completely with the new policy.
Third party monitors can also run these checks themselves to verify that the server has changed the policy correctly and honestly.
The security notion for this model includes the consistency notion described in
\cref{sec:simple-merkle-and-auditing} as well as the privacy property. Further analysis of the security of this model follows in \cref{sec:security_analysis}.

\ifarxiv
\section{Globally Client-Verifiable Authorization Records}
\label{sec:rsa-record-appendix}
Constructions of transparency dictionaries such as Verdict~\cite{verdict}  are efficient only when amortizing the cost of updates into ``epochs.''
For a package repository, we may not want to wait for a new epoch to allow downloads of a newly published package.
Noting that most packages are only updated once, and that \emph{any} registration of a new package is correct, we propose a two-tiered structure for an authorization record.
First, an append-only authenticated dictionary (AAD)~\cite{alin-thesis,kvac,txn20} (with no support for history/updates) for the initial package policies.
Second, a simple authenticated dictionary (SAD, like a Merkle binary prefix tree) storing the history.
Clients store the digests for both.
When fetching an update, clients get (1) the new digest for the AAD and a proof that policies have only been added, and (2) a list of updates for the SAD, along with membership proofs for the current policy for each updated entry.
The clients verify that no initial package policy has been modified, and can replay each update to the history dictionary to reconstruct its new digest, then verify that this digest is signed by the monitors in the consistency protocol.
When looking up a never-updated package, they require a lookup proof for the AAD and a non-membership proof for the SAD.
When looking up a package with updates, they just require the SAD lookup proof.
When a new package is registered, the server must (1) add it to the AAD and compute append-only proofs from previous digests, and (2) computes a proof of non-membership against the authenticated dictionary.
When package ownership changes, the server stores the update, the current policy, and a membership proof for the current policy; it then updates the policy in the SAD, and updates the non-membership proofs for all never-updated packages.

Concretely, this could be instantiated with the ``insert-only'' key-value accumulator KVaC~\cite{kvac} as the AAD and a Merkle BPT as the SAD.
We can use VeRSA~\cite{versa} to approximate the performance of KVaC, giving a baseline for a repository with $p=100,000$ packages and \num{10} updates per hour, of which only \num{1} changes an existing package.
On an initial download, a client downloads a KVaC digest (\SI{512}{\byte}) and a Merkle digest (\SI{32}{byte}); this is all they need to store.
Fetching and verifying a package's policy for a never-updated package takes an AAD lookup verification (\SI{20}{\milli\second} with a \SI{0.5}{\kibi\byte} proof) and a SAD nonmembership verification (\SI{10}{\micro\second} with a \SI{1.5}{\kibi\byte} proof).
For a package that \emph{has} been updated, a user verifies a SAD lookup proof (\SI{10}{\micro\second} with a \SI{1.5}{\kibi\byte} proof).
Each time a client updates their digest, they must fetch the new AAD digest and an append-only proof (\SI{100}{\milli\second} with a \SI{4}{\kibi\byte} proof).
To verify the correctness of each individual update, the client verifies a SAD membership proof (\SI{10}{\micro\second}, \SI{1.5}{\kibi\byte}), performs the same change to their own digest (\SI{50}{\micro\second}), and checks the correctness of the update.
Even a verifying a year's worth of updates should only take about a second and \SI{80}{\mebi\byte}.
To register a new package (about 10 per hour), the server must update the AAD lookup proofs for each of the $O(p)$ never-updated packages (\SI{100}{\second} total, but embarassingly parallelizable---a single machine could bring it to about \SI{10}{\second}.
The server must also compute AAD append-only proofs in amortized $O(\lg p)$ time (\SI{10}{\milli\second} on average, with spikes that can parallelize up to a factor of $O(\lg p)$).
Finally, the server must compute a SAD nonmembership proof for the package (\SI{500}{\nano\second})
To update a package's ownership (about 1 per hour), the server must update the SAD (\SI{50}{\micro\second}), update SAD nonmembership proofs for $O(p)$ packages (\SI{500}{\nano\second} per, about \SI{50}{\milli\second} total), and store the old policy along with the update.
\fi

\end{document}